\documentclass[%
 reprint,
 superscriptaddress,
 onecolumn,
nofootinbib,
 amsmath,amssymb,
 aps,
rmp,
longbibliography,
]{revtex4-2}

\usepackage{amsmath,amssymb, amsthm, mathtools}
\usepackage{url}
\usepackage{bm}
\usepackage{braket}
\usepackage{xcolor}
\definecolor{crimson}{RGB}{165,28,48}
\usepackage[colorlinks=true,allcolors=crimson]{hyperref}
\usepackage{microtype}

\usepackage{autonum}

\makeatletter
\let\cat@comma@active\@empty
\makeatother

\begin{document}

\title{Summary statistics of learning link changing neural representations to behavior}

\author{Jacob A. Zavatone-Veth}
\email{jzavatoneveth@fas.harvard.edu}
\affiliation{Center for Brain Science, Harvard University, Cambridge, MA, USA}
\affiliation{Society of Fellows, Harvard University, Cambridge, MA, USA}

\author{Blake Bordelon}
\email{blake\_bordelon@g.harvard.edu}
\affiliation{John A. Paulson School of Engineering and Applied Sciences, Harvard University, Cambridge, MA, USA}
\affiliation{Center for Brain Science, Harvard University, Cambridge, MA, USA}

\author{Cengiz Pehlevan}
\email{cpehlevan@seas.harvard.edu}

\affiliation{John A. Paulson School of Engineering and Applied Sciences, Harvard University, Cambridge, MA, USA}
\affiliation{Center for Brain Science, Harvard University, Cambridge, MA, USA}
\affiliation{Kempner Institute for the Study of Natural and Artificial Intelligence, Harvard University, Cambridge, MA, USA}

\begin{abstract}
    How can we make sense of large-scale recordings of neural activity across learning? Theories of neural network learning with their origins in statistical physics offer a potential answer: for a given task, there are often a small set of summary statistics that are sufficient to predict performance as the network learns. Here, we review recent advances in how summary statistics can be used to build theoretical understanding of neural network learning. We then argue for how this perspective can inform the analysis of neural data, enabling better understanding of learning in biological and artificial neural networks. 
\end{abstract}

\date{\today}

\maketitle

\section{Introduction}

Experience reshapes neural population activity, molding an animal's representations of the world as it learns to perform new tasks. Thanks to advances in experimental technologies, it is just now becoming possible to measure changes in the activity of large neural populations across the course of learning \citep{masset2022drifting,fink2025experience,kriegeskorte2021geometry,steinmetz2021neuropixels,zhong2025unsupervised,sun2025orthogonalized,vaidya2025expanding}. However, with this new capability comes the challenge of identifying which features of high-dimensional activity patterns are meaningful for understanding learning. While analyses of representations have begun how to elucidate how learning reshapes the structure of activity, it is not in general clear whether these measurements are sufficient to understand how representational changes relate to behavior \citep{krakauer2017behavior,sucholutsky2024aligned,kriegeskorte2008rsa,kriegeskorte2021geometry}.

In this Perspective, we propose that the principled identification of \textbf{summary statistics of learning} offers a possible path forward. This framework is grounded in theories of the statistical physics of learning in neural networks, which show that low-dimensional summary statistics are often sufficient to predict task performance over the course of learning \citep{watkin1993rule,engel2001statistical,zdeborova2016thresholds}. We argue that thinking systematically about summary statistics gives new insight into what existing approaches of quantifying neural representations reveal about learning, and allows identification of what additional measurements would be required to constrain models of plasticity. We emphasize that the goal of this Perspective is not to advocate for the use of a particular set of summary statistics, but rather to explain the general philosophy of this approach to understanding learning in high dimensions. 

\section{What is a summary statistic?}\label{sec:definition}

We posit that summary statistics of learning must satisfy two minimal desiderata: 
\begin{enumerate}
    \item \textbf{They must be low-dimensional.} That is, their dimension is low relative to the number of neurons in the network of interest. Indeed, most summary statistics we will encounter are determined by averages over the population of neurons. 

    \item \textbf{They must be sufficient to predict behavior across learning.} From a theoretical standpoint, there should exist a closed set of equations describing the evolution of the summary statistics that predict the network's performance. 
\end{enumerate}
As we will illustrate with concrete examples in Section \ref{sec:theory}, summary statistics satisfying these two desiderata are often highly interpretable thanks to their clear relationship to the network architecture and learning task. However, the summary statistics relevant for predicting performance may not be sufficient to predict all statistical properties of population activity. We will elaborate on this issue, and the resulting limitations of descriptions based on summary statistics alone, in Section \ref{sec:measurement}. 

Our use of the term ``summary statistics'' follows work by Ben Arous and colleagues \citep{benarous2022high,benarous2023eigenspaces}. In the literature on the statistical physics of learning, the quantities that we refer to as summary statistics are often termed ``order parameters'' \citep{mezard1987spin,watkin1993rule,engel2001statistical,zdeborova2016thresholds}. We prefer to use the former, more general term as it better captures the goal of these reduced descriptions in a neuroscientific context: we aim to summarize the features of neural activity relevant for learning. 

\section{Summary Statistics in Theories of Neural Network Learning}\label{sec:theory}

We now review how summary statistics emerge naturally in theoretical analyses of neural network learning. Out of many theoretical results, we focus on two example settings: online learning from high-dimensional data in shallow networks, and batch learning in wide and deep networks \citep{benarous2023eigenspaces,goldt2019dynamics,saad1995line,cui2023bayes,zv2021scale,bordelon2023self,zavatone2022contrasting,saxe2013exact,bordelon2025rnn,arnaboldi2023unifying,vanmeegen2025coding,watkin1993rule,engel2001statistical,zdeborova2016thresholds}. These model problems illustrate how relevant summary statistics may be identified given a task, network architecture, and learning rule. 

\subsection{Online learning in shallow neural networks with high dimensional data}

Classical models of online gradient descent learning in high dimensions can be often be summarized with simple summary statistics \citep{watkin1993rule,engel2001statistical, benarous2022high,arnaboldi2023unifying,goldt2019dynamics,goldt2020hidden,biehl1995online,saad1995line}. In this section, we discuss how  the generalization performance of perceptrons and shallow (two-layer) neural networks trained on large quantities of high dimensional data can be summarized by simple weight alignment measures. Most simply, the perceptron model $f(\mathbf{x}) = \sigma\left(\frac{1}{\sqrt{D}} \mathbf{w} \cdot \mathbf{x} \right)$ seeks to learn a weight vector $\mathbf w \in \mathbb{R}^D$ which correctly classifies a finite set of randomly sampled training input-output pairs $(\mathbf{x}_{\mu},y_{\mu})$. If the inputs are random, $\mathbf x_{\mu} \sim \mathcal{N}(0,\mathbf I_{D})$, and the targets $y_\mu = y(\mathbf{x}_{\mu})$ are generated by a \textbf{teacher network} $y(\mathbf{x}) = \sigma\left(  \frac{1}{\sqrt D} \mathbf w_\star \cdot \mathbf x  \right)$, then the generalization performance (performance of the model on new \textit{unseen data}, $\mathbb{E}_{\mathbf{x}}[(f(\mathbf{x})-y(\mathbf{x}))^2]$) is completely determined by the overlap of $\mathbf w$ with itself and with the target direction $\mathbf w_\star$
\begin{equation}
    Q = \frac{1}{D} \mathbf w \cdot \mathbf w \ , \ R = \frac{1}{D} \mathbf w \cdot \mathbf w_\star .
\end{equation}
If the learning rate is scaled appropriately with the dimension $D$, the high-dimensional (large-$D$) limit of online stochastic gradient descent is given by a deterministic set of equations for $Q$ and $R$: 
\begin{equation} \label{eqn:saadsolla}
    \frac{d}{d\tau} \begin{bmatrix}
        Q(\tau)
        \\
        R(\tau)
    \end{bmatrix} = \mathbf{F}[ Q(\tau), R(\tau) ] ,
\end{equation} 
where the continuous training `time' $\tau$ is the ratio of the number of samples seen to the dimension and $\mathbf F : \mathbb{R}^2 \to \mathbb{R}^2$ is a nonlinear function that depends on the learning rate, the loss function, and the link function $\sigma(\cdot)$ \citep{engel2001statistical, benarous2022high,arnaboldi2023unifying,goldt2019dynamics,saad1995line}. Integrating this update equation allows one to predict the evolution of the generalization error as more training data are provided to the algorithm. Despite the infinite dimensionality of the original optimization problem, only two dimensions are necessary to capture the dynamics of generalization error.

The analysis of online perceptron learning can be extended to two layer neural networks with a small number of hidden neurons $N$, 
\begin{align}
    f(\mathbf x) = \frac{1}{N} \sum_{i=1}^N a_i \  \phi\left( h_i(\mathbf x) \right)  \quad   h_i(\mathbf x) = \frac{1}{\sqrt D} \mathbf w_i \cdot \mathbf x \ , \ i \in \{1,..., N \} . 
    \\
    y(\mathbf x) = \sigma\left( h^\star_1(\mathbf x), ..., h^\star_K(\mathbf x) \right)   \quad  h^\star_k(\mathbf x) = \frac{1}{\sqrt D} \mathbf w^\star_k  \cdot  \mathbf x \ , \   k \in \{1,..., K \} .
\end{align}
In this setting with isotropic random data, the relevant summary statistics are the readout weights $\mathbf{a} \in \mathbb{R}^{N}$, along with \textbf{overlap matrices} $\mathbf Q \in \mathbb{R}^{N \times N}$ and $\mathbf R \in \mathbb{R}^{N \times K}$ with entries
\begin{equation}
    Q_{ij} = \frac{1}{D} \mathbf w_i \cdot \mathbf w_j \ , \ R_{ik} = \frac{1}{D} \mathbf w_i \cdot \mathbf w^{\star}_k 
\end{equation}
For this system, we can track the gradient descent dynamics for $\mathbf a$, $\mathbf{Q}$, and $\mathbf{R}$ through a generalization of Equation \eqref{eqn:saadsolla} \citep{goldt2019dynamics,saad1995line,biehl1995online,goldt2020hidden}. This reduces the dimensionality of the dynamics from the $N + DN$ trainable parameters $\{ a_i \}$, $\{ w_j \}$ to $N + N^2 + NK$ summary statistics, which is significant when $D \gg N+K$. This reduction enables the application of analyses that cannot scale to high dimensions, for instance control-theoretic methods to study optimal learning hyperparameters and curricula \citep{mori2025optimal,mignacco2025optimal}. Recent works have also begun to study approximations to these summary statistics when the network width $N$ is also large, as further dimensionality reduction if possible when $\mathbf Q$ and $\mathbf R$ have stereotyped structures \citep{montanari2025dynamical,arnaboldi2023unifying}.

Under what conditions is this reduction possible? Fundamentally, the summary statistics $\mathbf{a}$, $\mathbf{Q}$, and $\mathbf{R}$ are sufficient to determine the network's performance so long as the preactivations $h_{i}$ and $h_{k}^{\star}$ are approximately Gaussian. Thus, one can relax the assumption that the inputs $\mathbf{x}$ are exactly Gaussian so long as a central limit theorem applies to $h_{i}$ and $h_{k}^{\star}$ \citep{goldt2019dynamics,goldt2020hidden}. Moreover, one can allow for correlations between the different input dimensions so long as $h_{i}$ and $h_{k}^{\star}$ remain Gaussian. If $\mathbb{E}[\mathbf{x}\mathbf{x}^{\top}] = \mathbf{\Sigma}$, with a modification of the definition of the overlaps to $Q_{ij} = \frac{1}{D} \mathbf{w}_{i} \cdot \mathbf{\Sigma} \mathbf{w}_{j}$ and $R_{ik} = \frac{1}{D} \mathbf{w}_{i} \cdot \mathbf{\Sigma} \mathbf{w}^{\star}_{k}$ a similar reduction applies \citep{arnaboldi2023unifying}. One can even consider extensions to plasticity rules other than stochastic gradient descent. For example, online node perturbation leads to a different effective dynamics for the same set of summary statistics \citep{hara2011node,hara2013node}. 

How could the overlaps $\mathbf{Q}$ and $\mathbf{R}$ be accessed from measurements of neural activity? And, in the absence of detailed knowledge of a teacher network, how could one identify the relevant overlaps? Under the simple structural assumptions of these models, one could estimate the overlaps from covariances of network activity across stimuli, \textit{i.e.}, with isotropic inputs one has $\mathbb{E}_{\mathbf{x}}[ h_{i} h_{k}^{\star} ] = R_{ik}$ and $\mathbb{E}_{\mathbf{x}}[h_{i} h_{j}] = Q_{ij}$. Moreover, one can in some cases detect this underlying low-dimensional structure by examining the principal components of the learning trajectory \citep{benarous2023eigenspaces}. However, more theoretical work is required in this vein. 

\subsection{Learning in wide and deep neural networks}

\begin{figure}[t]
    \centering
    \includegraphics[width=6.9in]{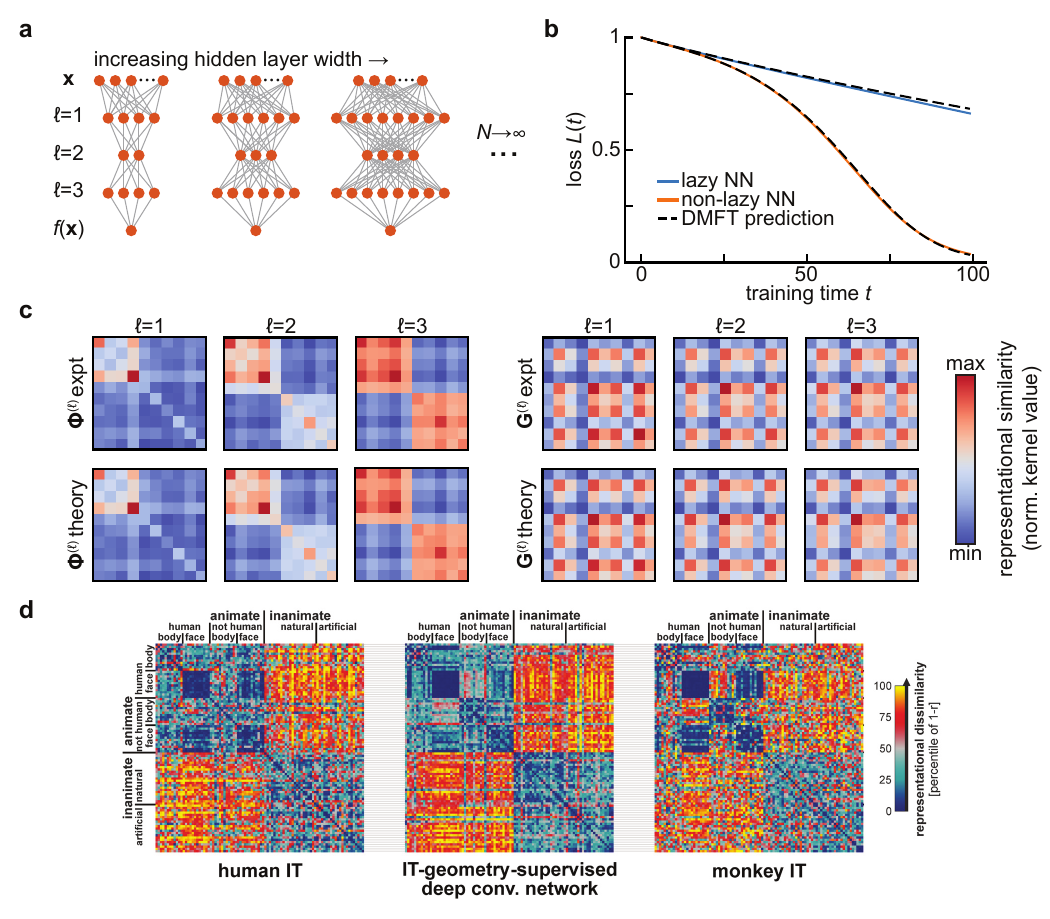}
    \caption{\textbf{Representational similarity kernels in wide neural network models and in the brain.} \\ \textbf{a}. Diagram of the infinite-width limit of a deep feedforward neural network. For a fixed input and output dimension, one considers a sequence of networks of increasing hidden layer widths. \textbf{b}. Predicting the performance of width-2500 fully-connected networks with three hidden layers and $\tanh$ activations over training using the dynamical mean-field theory described in Section \ref{sec:theory}. Networks are trained on a synthetic binary classification dataset of 10 examples, with 5 examples assigned each class at random. This leads to block structure in the final representations. Adapted from \citet{bordelon2023self}. \textbf{c}. The summary statistics in the dynamical mean field theory for the network in \textbf{b} are representational similarity kernels ($\mathbf{\Phi}^{(\ell)}$; \textit{left}) and gradient similarity kernels ($\mathbf{G}^{\ell}$; \textit{right}) for each layer. The top row shows kernels estimated from gradient descent training, and the bottom row the theoretical predictions. All kernels are shown at the end of training ($t=100$). Adapted from \citet{bordelon2023self}. \textbf{d}. Comparing representational similarity kernels across models and brains. Here, similarity is measured using the Pearson correlation $r$, and the \textit{dissimilarity} $1-r$ is plotted as a heatmap. Kernels resulting from fMRI measurements of human inferior temporal (IT) cortex (\textit{left}) and electrophysiological measurements of macaque monkey IT cortex (\textit{right}) are compared with the kernel for features from a deep convolutional neural network after optimal re-weighting to match human IT (\textit{center}). Adapted from Figure 10 of \citet{kaligh2014cortical} with permission from N. Kriegeskorte under a CC-BY License. }
    \label{fig:kernels}
\end{figure}

Another strategy to reduce the complexity of multilayer deep neural networks is to analyze the dynamics of learning in terms of representational similarity matrices (kernels) for each hidden layer of the network. Consider, for example, a deep fully-connected network with input $\mathbf{x} \in \mathbb{R}^{D}$,
\begin{equation}
\begin{split}
    f(\mathbf{x},t) &= \frac{1}{\gamma \sqrt{N}} \sum_{i=1}^N w_i(t) \phi(h^{(L)}_i(\mathbf{x},t)) , 
    \\   
    h^{(\ell+1)}_i(\mathbf{x},t) &= \frac{1}{\sqrt N} \sum_{j=1}^N W^{(\ell)}_{ij}(t) \phi(h^{(\ell)}_j(\mathbf{x},t)) , \quad \ell \in \{1,\ldots,L+1\},
    \\ 
    h^{(1)}_i(\mathbf{x},t) &= \frac{1}{\sqrt D} \sum_{j=1}^{D} W^{(0)}_{ij}(t) x_j,
\end{split}
\end{equation}
where $t$ denotes training time. Instead of using online stochastic gradient descent to train the weights as we did in the preceding section, suppose we use gradient flow to minimize the average error on a fixed set of training examples. Moreover, instead of considering a regime where the hidden layer width $N$ is small relative to the input dimension $D$, let us now consider very wide networks with $N \gg D$ (Figure \ref{fig:kernels}a). 

What are the relevant summary statistics in this case? Applying the chain rule to the dynamics of the network outputs, one finds the differential equation
\begin{equation}
    \frac{d}{dt} f(\mathbf{x},t) = - \mathbb{E}_{\mathbf{x}'} \sum_{\ell}  G^{(\ell+1)}(\mathbf{x},\mathbf{x}',t,t) \Phi^{(\ell)}(\mathbf{x},\mathbf{x}',t,t) \frac{\partial \mathcal L}{\partial f(\mathbf{x}',t)},
\end{equation}
where $\mathcal L$ is the loss function and $\mathbb{E}_{\mathbf{x}'}$ denotes expectation over the training dataset \citep{jacot2018neural,lee2019wide,bordelon2023self}. Here, 
\begin{equation}
    \Phi^{(\ell)}(\mathbf{x},\mathbf{x}',t,t') = \frac{1}{N} \sum_{i=1}^N \phi(h_i^{(\ell)}(\mathbf{x},t)) \phi(h_i^{(\ell)}(\mathbf{x}',t')) 
\end{equation}
are \textbf{representational similarity matrices}, and 
\begin{equation}
    G^{(\ell)}(\mathbf{x},\mathbf{x}',t,t') = \frac{1}{N} \sum_{i=1}^N g_i^{(\ell)}(\mathbf{x},t) g_i^{(\ell)}(\mathbf{x}',t') , \quad g^{(\ell)}_i(\mathbf{x},t) \equiv \gamma \sqrt N \frac{\partial f(\mathbf{x},t)}{\partial h^{(\ell)}_i(\mathbf{x},t)} ,
\end{equation}
are \textbf{gradient similarity matrices}, which respectively compare the hidden states $\phi(h^{(\ell)}_{i}(\mathbf x,t))$ and the gradient signals $g^{(\ell)}_i(\mathbf{x},t)$ at each hidden layer $\ell$ for each pair of data points $(\mathbf{x},\mathbf{x}')$ and each pair of training times $(t,t')$. Thus, as $\Phi^{(\ell)}$ and $G^{(\ell)}$ determine the dynamics of $f$, these matrices are suitable summary statistics of learning if they are low-dimensional relative to the set of synaptic weights, and if we can write down a closed set of equations for their dynamics. 

First, it is easy to see that the criterion of dimensionality reduction requires that the number of training examples $P$ is much less than the network width $N$, as the number of similarity matrix elements and the number of synaptic weights are of order $P^2$ and $N^2$, respectively. Second, it turns out that one can close the equations for $\Phi^{(\ell)}$ and $G^{(\ell)}$ provided that the width is large and that the synaptic weights start from an uninformed initial condition (\textit{i.e.}, Gaussian random matrices) \citep{jacot2018neural,lee2019wide,yang2021tensor, bordelon2023self}. Depending on how weights and learning rates are scaled, one can obtain different types of large-width ($N \to \infty$) limits (Figure \ref{fig:kernels}b). In the \textit{lazy / kernel} limit where $\gamma$ is constant, these representational similarity matrices are static over the course of learning \citep{jacot2018neural,lee2019wide}. However, an alternative scaling ($\gamma \propto \sqrt{N}$) can be adopted where these objects evolve in a task-dependent manner even as $N \to \infty$ (Figure \ref{fig:kernels}c) \citep{yang2021tensor, bordelon2023self}. 

While this provides a description of the training dynamics of a model under gradient flow, one can extend this description in terms of similarity matrices to other learning rules which use approximations of the backward pass variables $\tilde{g}^{(\ell)}_i(\mathbf{x},t)$, which we called pseudo-gradients in \citet{bordelon2023influence}. Such rules include Hebbian learning, feedback alignment, and direct feedback alignment \citep{hebb2005organization, lillicrap2016random, nokland2016direct}. In this case, the relevant summary statistics to characterize the prediction dynamics of the network include the gradient-pseudogradient correlation, which measures the alignment between the gradients used by the learning rule and the gradients that one would have used with gradient flow,
\begin{equation}\label{eq:influence_lr_tilde_G}
    \tilde{G}^{(\ell)}(\mathbf{x},\mathbf{x}',t,t') = \frac{1}{N} \sum_{i=1}^N g^{(\ell)}_i(\mathbf{x},t) \tilde{g}^{(\ell)}_i(\mathbf{x}',t'), 
\end{equation}
as $\tilde{G}^{(\ell)}$ governs the evolution of the function output:
\begin{equation}
     \frac{d}{dt} f(\mathbf{x},t) = - \mathbb{E}_{\mathbf{x}'} \sum_{\ell}  \tilde G^{(\ell+1)}(\mathbf{x},\mathbf{x}',t,t) \Phi^{(\ell)}(\mathbf{x},\mathbf{x}',t,t) \frac{\partial \mathcal L}{\partial f(\mathbf{x}',t)}. 
\end{equation}

\clearpage

\section{Implications for Neural Measurements}\label{sec:measurement}

The two example settings detailed in Section \ref{sec:theory} show how the relevant summary statistics of learning depend on network architecture and learning rule. Theoretical studies are just beginning to map out the full space of possible summary statistics for different network architectures \citep{benarous2023eigenspaces,goldt2019dynamics,saad1995line,cui2023bayes,zv2021scale,bordelon2023self,zavatone2022contrasting,saxe2013exact,bordelon2025rnn,arnaboldi2023unifying,vanmeegen2025coding,engel2001statistical,zdeborova2016thresholds}. Though details of the relevant summary statistics vary depending on the scaling regime and task---as illustrated by the examples above, where network width, training dataset size, and learning rule change the relevant statistics and their effective dynamics---they share broad structural principles. In all cases, summary statistics are defined by (weighted) averages over sub-populations of neurons within the network of interest, \textit{e.g.}, correlations of activity with task-relevant variables, or autocorrelations of activity within a particular layer in a deep network. Thanks to these common structural features, these varied theories of summary statistics have common implications for the analysis and interpretation of neuroscience experiments.

\subsection{Benign subsampling}

The summary statistics encountered in Section \ref{sec:theory} are robust to subsampling thanks to their basic nature as averages over the population of neurons. These statistical theories in fact post a far stronger notion of benign subsampling: they result in neurons that are statistically exchangeable. This is highly advantageous from the perspective of long-term recordings of neural activity, as reliable measurement of summary statistics does not require one to track the exact same neurons over time. Instead, it suffices to measure a sufficiently large subpopulation on any given day. This obviates many of the challenges presented by tracking neurons over multiple recording sessions \citep{masset2022drifting}. Moreover, the variability and bias introduced by estimating summary statistics from a limited subset of relevant neurons can be characterized systematically \citep{kang2025spectral,bordelon2024finite,zv2022asymptotics}. Taken together, these properties mean that summary statistics are relatively easy to estimate given limited neural measurements, provided that exchangability is not too strongly violated \citep{gao2017theory}. We will return to this question in the Discussion, as a detailed analysis of the effects of non-identical neurons will be an important topic for future theoretical work. There are limits, however, to how far one can subsample. For instance, representational similarity kernels are more affected by small, coordinated changes in the tuning of many neurons than large changes in single-neuron tuning (Figure \ref{fig:drift}) \citep{kriegeskorte2021geometry}. Determining the minimum number of neurons one must record in order to predict generalization dynamics across learning will be an important subject for future theoretical work \citep{gao2017theory,kriegeskorte2021geometry}.

\subsection{Invariances and representational drift}

Though by our definition the summary statistics mentioned in Section \ref{sec:theory} are sufficient to predict the network's performance, they are not sufficient statistics for all properties of the neural code. In particular, in part because they arise from theories in which neurons become exchangable, they have many invariances. These invariances mean that individual tuning curves can change substantially without altering the population-level computation \citep{kriegeskorte2021geometry}. For instance, the representational similarity kernels are invariant under rotation of the neural code at each layer, enabling complete reorganization of the single-neuron code without any effect on behavior. Similarly, overlaps with task-relevant directions are invariant to changes in the null space of those low-dimensional projections. These invariances mean that focusing on summary statistics of learning sets a particular aperture on what aspects of representations one can assay.

At the same time, the invariances of summary statistics have important consequences for functional robustness. In particular, they are closely related to theories of representational drift, the seemingly puzzling phenomenon of continuing changes in neural representations of task-relevant variables despite stable behavioral performance \citep{rule2019causes,masset2022drifting}. Many models of drift explicitly propose that representational changes are structured in such a way that certain summary statistics are preserved (Figure \ref{fig:drift}a) \citep{masset2022drifting,pashakhanloo2023drift,qin2023coordinated}. Identifying the invariances of the summary statistics sufficient to determine task performance can allow for a more systematic characterization of what forms of drift can be accommodated by a given network. Conversely, identifying the invariances of a representation once task performance stabilizes might suggest which summary statistics are relevant for the learning problem at hand. 

\subsection{Universality}

An important lesson from the theory of high-dimensional statistics is that of \emph{universality}: certain coarse-grained statistics are asymptotically insensitive to the details of the distribution. The most prominent example of statistical universality is the familiar central limit theorem: the distribution of the sample mean of independent random variables tends to a Gaussian as the number of samples becomes large. A broader class of universality principles arise in random matrix theory: the distribution of eigenvalues and eigenvectors of a random matrix often become insensitive to details of the distribution of the elements as the matrix becomes large. Most famously, the Mar\v{c}enko-Pastur theorem specifies that the singular values of a matrix with independent elements have a distribution that depends only on the mean and variance of the elements \citep{marchenko1967distribution}. In the context of learning problems, universality manifests through insensitivity of the model performance to details of the distributions of parameters or of features \citep{hu2022universality,misiakiewicz2024non}. 

\begin{figure}[ht!]
    \centering
    \includegraphics[width=6.8in]{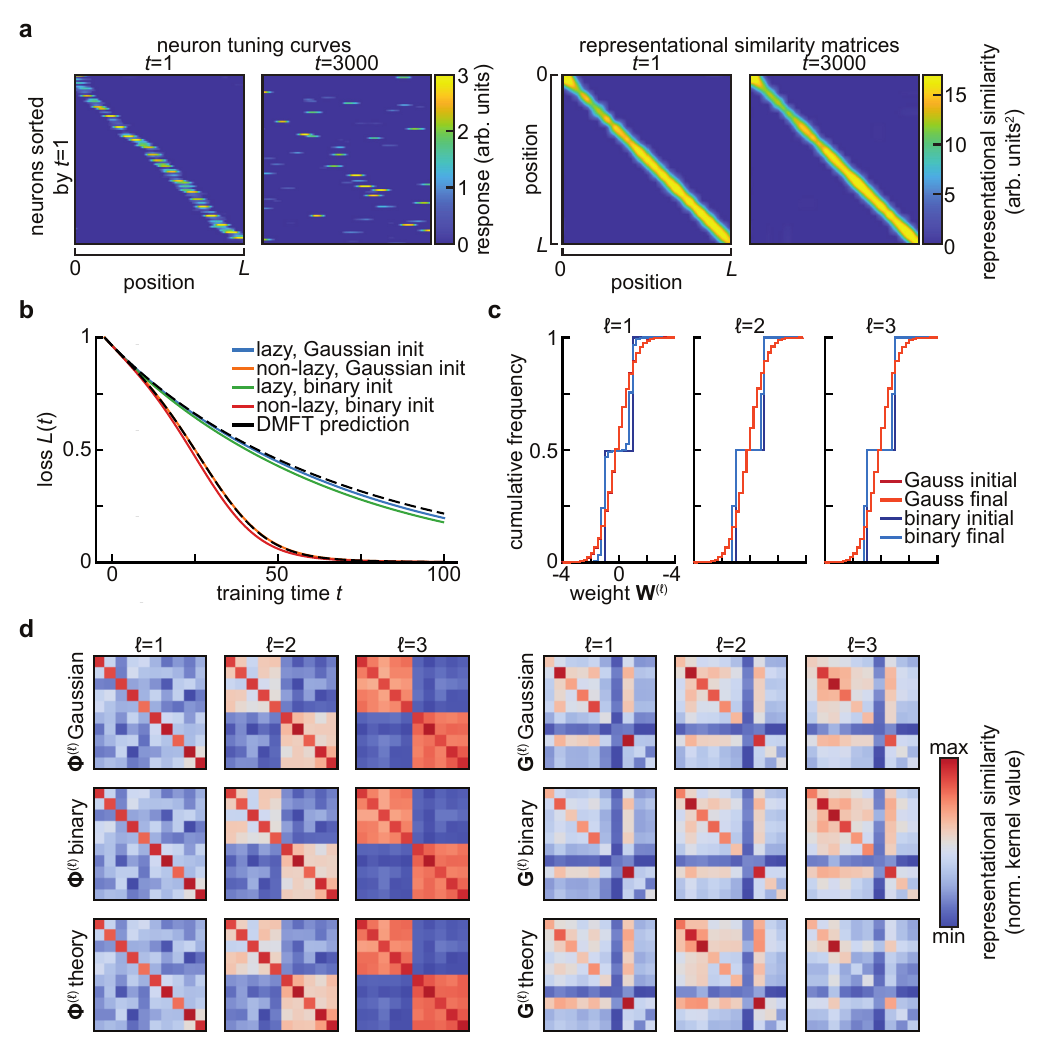}
    \caption{\textbf{Invariance and universality in summary statistics.} \\ \textbf{a}. Stable summary statistics despite drifting single-neuron responses. In \citet{qin2023coordinated}'s model of representational drift, single neurons are strongly tuned to a spatial variable, yet their tuning changes dramatically over time (\textit{left}). Despite this drift, the similarity of the population representations of different spatial positions remains nearly constant (\textit{right}). Adapted from Figure 5e of \citet{qin2023coordinated}, of which C.P. is the corresponding author. \textbf{b}. Universality of summary statistics in wide and deep networks with respect to the distribution of initial weights. Setting is as in Figure \ref{fig:kernels}b-c, but also including a network for which the weights are initially drawn from $\{-1,+1\}$ with equal probability. Here, $N = 2000$, and a different realization of the random task is sampled relative to Figure \ref{fig:kernels}b-c, so the loss curves are not identical. \textbf{c}. Cumulative distribution of weights at the start (\textit{initial}) and end (\textit{final}) of training for the networks shown in (\textbf{b}). Note that the small change in the weight distributions for the Gaussian-initialized networks is not visible at this resolution, and that one expects the size of weight changes to scale with $1/\sqrt{N}$ \citep{bordelon2023self}. \textbf{d}. Feature and gradient kernels at the end of training for the networks in \textbf{b}. No substantial differences are visible between networks initialized with different weight distributions. }
    \label{fig:drift}
\end{figure}

From the perspective of summary statistics, statistical universality can allow simple theories to make informative macroscopic predictions even if they do not capture detailed properties of single neurons. For instance, the mean-field description of the learning dynamics of wide neural networks introduced in Section \ref{sec:theory} are universal in that they depend on the initial distribution of hidden layer weights only through its mean and variance, even though the details of that distribution will affect the distribution of weights throughout training (Figure \ref{fig:drift}b-d) \citep{golikov2022nongaussian,williams1996infinite}. Like the invariances to transformations of the neural population code mentioned before, this is nonetheless a double-edged sword: these universality properties mean that focusing on predicting performance commits one to coarse-graining away certain microscopic aspects of neural activity. Though these features are not required to predict macroscopic behavior, they may be important for understanding biological mechanisms.

\clearpage

\clearpage

\section{Discussion}

The core insight of the statistical mechanics of learning in neural networks is the existence of low-dimensional summary statistics sufficient to predict behavioral performance. We have reviewed how different summary statistics emerge depending on network architecture and task, how summary statistics might be estimated from experimental recordings, and what this perspective reveals about existing approaches to quantifying representational changes over learning. We now conclude by discussing complementary summary statistics of neural representations that arise from alternative desiderata, and future directions for theoretical inquiry. 

A significant line of recent work in neuroscience aims to quantify neural representations and compare them across networks through analysis of representational similarity matrices $\Phi^{(\ell)}(\mathbf{x},\mathbf{x}')$ \citep{kriegeskorte2008rsa,sucholutsky2024aligned,williams2021shape,williams2024equivalence}. In Section \ref{sec:theory}, we have seen how these kernel matrices arise naturally as summary statistics of forward signal propagation in wide and deep neural networks (Figure \ref{fig:kernels}c-d). At the same time, those results show that tracking only feature kernels is not in general sufficient to predict performance over the course of learning. One needs access also to coarse-grained information about the plasticity rule in the form of gradient kernels (either $G^{(\ell)}$ or $\tilde G^{(\ell)}$), and to information about the network outputs (for instance ${\partial \mathcal L}/{\partial f}$). More theoretical work is required to determine how to reliably estimate these gradient kernels from data, thereby providing a means to gain coarse-grained information about the underlying plasticity rule. % Suggestively, however, the above equations suggest that one might be able to estimate the effective gradient kernel $\tilde{G}^{(\ell)}(x,x')$ if one can track changes in function outputs $f(x)$ and feature kernels $\Phi^{(\ell)}(x,x')$ over time, which are both experimentally accessible. That is, one could try to solve equation \eqref{eq:influence_lr_tilde_G} for an estimate of the effective gradient kernel $\tilde{G}$, and then compare that estimate to the kernel resulting from a particular model of plasticity \citep{bordelon2023influence}. We emphasize that this idea is rather speculative, as the equation for $\tilde{G}$ is underdetermined. 

The summary statistics discussed here explicitly depend on the architecture and nature of plasticity in the neural network of interest, as they seek to predict its performance over learning. A distinct set of summary statistics arises if one aims to study what features of a representation are relevant for an \textit{independently-trained} decoder. In this line of work, one regards the representation as fixed, rather than considering end-to-end training of the full network as we considered here. If the decoder is a simple linear regressor that predicts a continuous variable, the relevant summary statistics of the representation are just its mean and covariance across stimuli \citep{hu2022universality,misiakiewicz2024non}. Given a particular task, the covariance can be further distilled into the rate of decay of its eigenvalues and of the projections of the task direction into its eigenvectors \citep{hastie2022surprises,bordelon2022population,canatar2021spectral,canatar2024spectral,atanasov2024scaling,williams2024equivalence,harvey2024what,bordelon2023td}. For categorically-structured stimuli, a substantial body of work has elucidated the summary statistics that emerge from assuming that one wants to divide the data according to a random dichotomy \citep{chung2018classification,cohen2020separability,bernardi2020geometry,farrell2022capacity,engel2001statistical,zv2022capacity,sorscher2022fewshot,harvey2024what}. 

The models reviewed here are composed of exchangable neurons, which simplifies the relevant summary statistics and renders them particularly robust to subsampling. However, the brain has rich structure that can affect which summary statistics are sufficient to track learning and how those summary statistics may be measured. Biological neural networks are embedded in space, and their connectivity and selectivity is shaped by spatial structure \citep{khona2025modules,chklovskii2002wiring,stiso2018spatial}. Notably, many sensory areas are topographically organized: neurons with similar response properties are spatially proximal \citep{kandler2009tonotopic,murthy2011maps}. Moreover, neurons can be classified into genetically-identifiable cell types \citep{zhang2023atlas}, which may play distinct functional roles during learning \citep{hirokawa2019frontal,fink2025experience}. Future theoretical work must contend with these biological complexities in order to determine the relevant summary statistics of learning subject to these constraints.

\section*{Acknowledgements}

We are indebted to Nikolaus Kriegeskorte for sharing Figure 10 of \citet{kaligh2014cortical}, from which our Figure \ref{fig:kernels}d is derived. We thank Paul Masset, Venkatesh Murthy, Farhad Pashakhanloo, and Ningjing Xia for helpful discussions and comments on previous versions of this manuscript. 

J.A.Z.-V. is supported by the Office of the Director of the National Institutes of Health under Award Number DP5OD037354. The content is solely the responsibility of the authors and does not necessarily represent the official views of the National Institutes of Health. JAZV is further supported by a Junior Fellowship from the Harvard Society of Fellows. B.B. is supported by a Google PhD Fellowship. C.P. is supported by NSF grant DMS-2134157, NSF CAREER Award IIS-2239780, DARPA grant DIAL-FP-038, a Sloan Research Fellowship, and The William F. Milton Fund from Harvard University. This work has been made possible in part by a gift from the Chan Zuckerberg Initiative Foundation to establish the Kempner Institute for the Study of Natural and Artificial Intelligence.

\clearpage

\section*{Author Contributions}

Conceptualization, J.A.Z.-V., B.B., C.P.; Writing – Original Draft, J.A.Z.-V., B.B.; Visualization, J.A.Z.-V.; Writing – Review \& Editing, J.A.Z.-V., B.B., C.P.; Funding Acquisition, J.A.Z.-V. and C.P.

\bibliography{refs}

%apsrmp4-2.bst 2018-12-27 (MD) hand-edited version of apsrmp4-1.bst
%Control: author (3) reversed first dotless
\begin{thebibliography}{75}%
\makeatletter
\providecommand \@ifxundefined [1]{%
 \@ifx{#1\undefined}
}%
\providecommand \@ifnum [1]{%
 \ifnum #1\expandafter \@firstoftwo
 \else \expandafter \@secondoftwo
 \fi
}%
\providecommand \@ifx [1]{%
 \ifx #1\expandafter \@firstoftwo
 \else \expandafter \@secondoftwo
 \fi
}%
\providecommand \natexlab [1]{#1}%
\providecommand \enquote  [1]{``#1''}%
\providecommand \bibnamefont  [1]{#1}%
\providecommand \bibfnamefont [1]{#1}%
\providecommand \citenamefont [1]{#1}%
\providecommand \href@noop [0]{\@secondoftwo}%
\providecommand \href [0]{\begingroup \@sanitize@url \@href}%
\providecommand \@href[1]{\@@startlink{#1}\@@href}%
\providecommand \@@href[1]{\endgroup#1\@@endlink}%
\providecommand \@sanitize@url [0]{\catcode `\\12\catcode `\$12\catcode
  `\&12\catcode `\#12\catcode `\^12\catcode `\_12\catcode `\%12\relax}%
\providecommand \@@startlink[1]{}%
\providecommand \@@endlink[0]{}%
\providecommand \url  [0]{\begingroup\@sanitize@url \@url }%
\providecommand \@url [1]{\endgroup\@href {#1}{\urlprefix }}%
\providecommand \urlprefix  [0]{URL }%
\providecommand \Eprint [0]{\href }%
\providecommand \doibase [0]{https://doi.org/}%
\providecommand \selectlanguage [0]{\@gobble}%
\providecommand \bibinfo  [0]{\@secondoftwo}%
\providecommand \bibfield  [0]{\@secondoftwo}%
\providecommand \translation [1]{[#1]}%
\providecommand \BibitemOpen [0]{}%
\providecommand \bibitemStop [0]{}%
\providecommand \bibitemNoStop [0]{.\EOS\space}%
\providecommand \EOS [0]{\spacefactor3000\relax}%
\providecommand \BibitemShut  [1]{\csname bibitem#1\endcsname}%
\let\auto@bib@innerbib\@empty
%</preamble>
\bibitem [{\citenamefont {Arnaboldi}\ \emph {et~al.}(2023)\citenamefont
  {Arnaboldi}, \citenamefont {Stephan}, \citenamefont {Krzakala},\ and\
  \citenamefont {Loureiro}}]{arnaboldi2023unifying}%
  \BibitemOpen
  \bibfield  {author} {\bibinfo {author} {\bibnamefont {Arnaboldi},
  \bibfnamefont {Luca}}, \bibinfo {author} {\bibfnamefont {Ludovic}\
  \bibnamefont {Stephan}}, \bibinfo {author} {\bibfnamefont {Florent}\
  \bibnamefont {Krzakala}}, and\ \bibinfo {author} {\bibfnamefont {Bruno}\
  \bibnamefont {Loureiro}}} (\bibinfo {year} {2023}),\ \bibfield  {title}
  {\enquote {\bibinfo {title} {From high-dimensional \& mean-field dynamics to
  dimensionless {ODE}s: A unifying approach to {SGD} in two-layers networks},}\
  }in\ \href {https://proceedings.mlr.press/v195/arnaboldi23a.html} {\emph
  {\bibinfo {booktitle} {Proceedings of Thirty Sixth Conference on Learning
  Theory}}},\ \bibinfo {series} {Proceedings of Machine Learning Research},
  Vol.\ \bibinfo {volume} {195},\ \bibinfo {editor} {edited by\ \bibinfo
  {editor} {\bibfnamefont {Gergely}\ \bibnamefont {Neu}}\ and\ \bibinfo
  {editor} {\bibfnamefont {Lorenzo}\ \bibnamefont {Rosasco}}}\ (\bibinfo
  {publisher} {PMLR})\ pp.\ \bibinfo {pages} {1199--1227}\BibitemShut {NoStop}%
\bibitem [{\citenamefont {Atanasov}\ \emph {et~al.}(2024)\citenamefont
  {Atanasov}, \citenamefont {Zavatone-Veth},\ and\ \citenamefont
  {Pehlevan}}]{atanasov2024scaling}%
  \BibitemOpen
  \bibfield  {author} {\bibinfo {author} {\bibnamefont {Atanasov},
  \bibfnamefont {Alexander}}, \bibinfo {author} {\bibfnamefont {Jacob~A}\
  \bibnamefont {Zavatone-Veth}}, and\ \bibinfo {author} {\bibfnamefont
  {Cengiz}\ \bibnamefont {Pehlevan}}} (\bibinfo {year} {2024}),\ \bibfield
  {title} {\enquote {\bibinfo {title} {Scaling and renormalization in
  high-dimensional regression},}\ }\href@noop {} {\bibinfo  {journal} {arXiv
  preprint arXiv:2405.00592}\ }\BibitemShut {NoStop}%
\bibitem [{\citenamefont {Ben~Arous}\ \emph {et~al.}(2023)\citenamefont
  {Ben~Arous}, \citenamefont {Gheissari}, \citenamefont {Huang},\ and\
  \citenamefont {Jagannath}}]{benarous2023eigenspaces}%
  \BibitemOpen
\bibfield  {journal} {  }\bibfield  {author} {\bibinfo {author} {\bibnamefont
  {Ben~Arous}, \bibfnamefont {G{\'e}rard}}, \bibinfo {author} {\bibfnamefont
  {Reza}\ \bibnamefont {Gheissari}}, \bibinfo {author} {\bibfnamefont
  {Jiaoyang}\ \bibnamefont {Huang}}, and\ \bibinfo {author} {\bibfnamefont
  {Aukosh}\ \bibnamefont {Jagannath}}} (\bibinfo {year} {2023}),\ \bibfield
  {title} {\enquote {\bibinfo {title} {High-dimensional {SGD} aligns with
  emerging outlier eigenspaces},}\ }\href {https://arxiv.org/abs/2310.03010}
  {\bibfield  {journal} {\bibinfo  {journal} {arXiv}\ }}\Eprint
  {https://arxiv.org/abs/2310.03010} {arXiv:2310.03010 [cs.LG]} \BibitemShut
  {NoStop}%
\bibitem [{\citenamefont {Ben~Arous}\ \emph {et~al.}(2022)\citenamefont
  {Ben~Arous}, \citenamefont {Gheissari},\ and\ \citenamefont
  {Jagannath}}]{benarous2022high}%
  \BibitemOpen
  \bibfield  {author} {\bibinfo {author} {\bibnamefont {Ben~Arous},
  \bibfnamefont {G{\'e}rard}}, \bibinfo {author} {\bibfnamefont {Reza}\
  \bibnamefont {Gheissari}}, and\ \bibinfo {author} {\bibfnamefont {Aukosh}\
  \bibnamefont {Jagannath}}} (\bibinfo {year} {2022}),\ \bibfield  {title}
  {\enquote {\bibinfo {title} {High-dimensional limit theorems for {SGD}:
  Effective dynamics and critical scaling},}\ }in\ \href
  {https://proceedings.neurips.cc/paper_files/paper/2022/file/a224ff18cc99a71751aa2b79118604da-Paper-Conference.pdf}
  {\emph {\bibinfo {booktitle} {Advances in Neural Information Processing
  Systems}}},\ Vol.~\bibinfo {volume} {35},\ \bibinfo {editor} {edited by\
  \bibinfo {editor} {\bibfnamefont {S.}~\bibnamefont {Koyejo}}, \bibinfo
  {editor} {\bibfnamefont {S.}~\bibnamefont {Mohamed}}, \bibinfo {editor}
  {\bibfnamefont {A.}~\bibnamefont {Agarwal}}, \bibinfo {editor} {\bibfnamefont
  {D.}~\bibnamefont {Belgrave}}, \bibinfo {editor} {\bibfnamefont
  {K.}~\bibnamefont {Cho}}, \ and\ \bibinfo {editor} {\bibfnamefont
  {A.}~\bibnamefont {Oh}}}\ (\bibinfo  {publisher} {Curran Associates, Inc.})\
  pp.\ \bibinfo {pages} {25349--25362}\BibitemShut {NoStop}%
\bibitem [{\citenamefont {Bernardi}\ \emph {et~al.}(2020)\citenamefont
  {Bernardi}, \citenamefont {Benna}, \citenamefont {Rigotti}, \citenamefont
  {Munuera}, \citenamefont {Fusi},\ and\ \citenamefont
  {Salzman}}]{bernardi2020geometry}%
  \BibitemOpen
  \bibfield  {author} {\bibinfo {author} {\bibnamefont {Bernardi},
  \bibfnamefont {Silvia}}, \bibinfo {author} {\bibfnamefont {Marcus~K.}\
  \bibnamefont {Benna}}, \bibinfo {author} {\bibfnamefont {Mattia}\
  \bibnamefont {Rigotti}}, \bibinfo {author} {\bibfnamefont {J{\'e}r{\^o}me}\
  \bibnamefont {Munuera}}, \bibinfo {author} {\bibfnamefont {Stefano}\
  \bibnamefont {Fusi}}, and\ \bibinfo {author} {\bibfnamefont {C.~Daniel}\
  \bibnamefont {Salzman}}} (\bibinfo {year} {2020}),\ \bibfield  {title}
  {\enquote {\bibinfo {title} {The geometry of abstraction in the hippocampus
  and prefrontal cortex},}\ }\href {https://doi.org/10.1016/j.cell.2020.09.031}
  {\bibfield  {journal} {\bibinfo  {journal} {Cell}\ }\textbf {\bibinfo
  {volume} {183}}~(\bibinfo {number} {4}),\ \bibinfo {pages}
  {954--967.e21}}\BibitemShut {NoStop}%
\bibitem [{\citenamefont {Biehl}\ and\ \citenamefont
  {Schwarze}(1995)}]{biehl1995online}%
  \BibitemOpen
  \bibfield  {author} {\bibinfo {author} {\bibnamefont {Biehl}, \bibfnamefont
  {M}}, and\ \bibinfo {author} {\bibfnamefont {H}~\bibnamefont {Schwarze}}}
  (\bibinfo {year} {1995}),\ \bibfield  {title} {\enquote {\bibinfo {title}
  {Learning by on-line gradient descent},}\ }\href
  {https://doi.org/10.1088/0305-4470/28/3/018} {\bibfield  {journal} {\bibinfo
  {journal} {Journal of Physics A: Mathematical and General}\ }\textbf
  {\bibinfo {volume} {28}}~(\bibinfo {number} {3}),\ \bibinfo {pages}
  {643}}\BibitemShut {NoStop}%
\bibitem [{\citenamefont {Bordelon}\ \emph {et~al.}(2025)\citenamefont
  {Bordelon}, \citenamefont {Cotler}, \citenamefont {Pehlevan},\ and\
  \citenamefont {Zavatone-Veth}}]{bordelon2025rnn}%
  \BibitemOpen
  \bibfield  {author} {\bibinfo {author} {\bibnamefont {Bordelon},
  \bibfnamefont {Blake}}, \bibinfo {author} {\bibfnamefont {Jordan}\
  \bibnamefont {Cotler}}, \bibinfo {author} {\bibfnamefont {Cengiz}\
  \bibnamefont {Pehlevan}}, and\ \bibinfo {author} {\bibfnamefont {Jacob~A.}\
  \bibnamefont {Zavatone-Veth}}} (\bibinfo {year} {2025}),\ \bibfield  {title}
  {\enquote {\bibinfo {title} {Dynamically learning to integrate in recurrent
  neural networks},}\ }\href {https://arxiv.org/abs/2503.18754} {\bibfield
  {journal} {\bibinfo  {journal} {arXiv}\ }}\Eprint
  {https://arxiv.org/abs/2503.18754} {arXiv:2503.18754 [q-bio.NC]} \BibitemShut
  {NoStop}%
\bibitem [{\citenamefont {Bordelon}\ \emph {et~al.}(2023)\citenamefont
  {Bordelon}, \citenamefont {Masset}, \citenamefont {Kuo},\ and\ \citenamefont
  {Pehlevan}}]{bordelon2023td}%
  \BibitemOpen
  \bibfield  {author} {\bibinfo {author} {\bibnamefont {Bordelon},
  \bibfnamefont {Blake}}, \bibinfo {author} {\bibfnamefont {Paul}\ \bibnamefont
  {Masset}}, \bibinfo {author} {\bibfnamefont {Henry}\ \bibnamefont {Kuo}},
  and\ \bibinfo {author} {\bibfnamefont {Cengiz}\ \bibnamefont {Pehlevan}}}
  (\bibinfo {year} {2023}),\ \bibfield  {title} {\enquote {\bibinfo {title}
  {Loss dynamics of temporal difference reinforcement learning},}\ }in\ \href
  {https://proceedings.neurips.cc/paper_files/paper/2023/file/2ea04b568a2deb2d000c59f3a72829b5-Paper-Conference.pdf}
  {\emph {\bibinfo {booktitle} {Advances in Neural Information Processing
  Systems}}},\ Vol.~\bibinfo {volume} {36},\ \bibinfo {editor} {edited by\
  \bibinfo {editor} {\bibfnamefont {A.}~\bibnamefont {Oh}}, \bibinfo {editor}
  {\bibfnamefont {T.}~\bibnamefont {Naumann}}, \bibinfo {editor} {\bibfnamefont
  {A.}~\bibnamefont {Globerson}}, \bibinfo {editor} {\bibfnamefont
  {K.}~\bibnamefont {Saenko}}, \bibinfo {editor} {\bibfnamefont
  {M.}~\bibnamefont {Hardt}}, \ and\ \bibinfo {editor} {\bibfnamefont
  {S.}~\bibnamefont {Levine}}}\ (\bibinfo  {publisher} {Curran Associates,
  Inc.})\ pp.\ \bibinfo {pages} {14469--14496}\BibitemShut {NoStop}%
\bibitem [{\citenamefont {Bordelon}\ and\ \citenamefont
  {Pehlevan}(2022)}]{bordelon2022population}%
  \BibitemOpen
  \bibfield  {author} {\bibinfo {author} {\bibnamefont {Bordelon},
  \bibfnamefont {Blake}}, and\ \bibinfo {author} {\bibfnamefont {Cengiz}\
  \bibnamefont {Pehlevan}}} (\bibinfo {year} {2022}),\ \bibfield  {title}
  {\enquote {\bibinfo {title} {Population codes enable learning from few
  examples by shaping inductive bias},}\ }\href
  {https://doi.org/10.7554/eLife.78606} {\bibfield  {journal} {\bibinfo
  {journal} {eLife}\ }\textbf {\bibinfo {volume} {11}},\ \bibinfo {pages}
  {e78606}}\BibitemShut {NoStop}%
\bibitem [{\citenamefont {Bordelon}\ and\ \citenamefont
  {Pehlevan}(2023{\natexlab{a}})}]{bordelon2023influence}%
  \BibitemOpen
  \bibfield  {author} {\bibinfo {author} {\bibnamefont {Bordelon},
  \bibfnamefont {Blake}}, and\ \bibinfo {author} {\bibfnamefont {Cengiz}\
  \bibnamefont {Pehlevan}}} (\bibinfo {year} {2023}{\natexlab{a}}),\ \bibfield
  {title} {\enquote {\bibinfo {title} {The influence of learning rule on
  representation dynamics in wide neural networks},}\ }in\ \href@noop {} {\emph
  {\bibinfo {booktitle} {The Eleventh International Conference on Learning
  Representations}}}\BibitemShut {NoStop}%
\bibitem [{\citenamefont {Bordelon}\ and\ \citenamefont
  {Pehlevan}(2023{\natexlab{b}})}]{bordelon2023self}%
  \BibitemOpen
  \bibfield  {author} {\bibinfo {author} {\bibnamefont {Bordelon},
  \bibfnamefont {Blake}}, and\ \bibinfo {author} {\bibfnamefont {Cengiz}\
  \bibnamefont {Pehlevan}}} (\bibinfo {year} {2023}{\natexlab{b}}),\ \bibfield
  {title} {\enquote {\bibinfo {title} {Self-consistent dynamical field theory
  of kernel evolution in wide neural networks},}\ }\href@noop {} {\bibfield
  {journal} {\bibinfo  {journal} {Journal of Statistical Mechanics: Theory and
  Experiment}\ }\textbf {\bibinfo {volume} {2023}}~(\bibinfo {number} {11}),\
  \bibinfo {pages} {114009}}\BibitemShut {NoStop}%
\bibitem [{\citenamefont {Bordelon}\ and\ \citenamefont
  {Pehlevan}(2024)}]{bordelon2024finite}%
  \BibitemOpen
  \bibfield  {author} {\bibinfo {author} {\bibnamefont {Bordelon},
  \bibfnamefont {Blake}}, and\ \bibinfo {author} {\bibfnamefont {Cengiz}\
  \bibnamefont {Pehlevan}}} (\bibinfo {year} {2024}),\ \bibfield  {title}
  {\enquote {\bibinfo {title} {Dynamics of finite width kernel and prediction
  fluctuations in mean field neural networks},}\ }\href
  {https://doi.org/10.1088/1742-5468/ad642b} {\bibfield  {journal} {\bibinfo
  {journal} {Journal of Statistical Mechanics: Theory and Experiment}\ }\textbf
  {\bibinfo {volume} {2024}}~(\bibinfo {number} {10}),\ \bibinfo {pages}
  {104021}}\BibitemShut {NoStop}%
\bibitem [{\citenamefont {Canatar}\ \emph {et~al.}(2021)\citenamefont
  {Canatar}, \citenamefont {Bordelon},\ and\ \citenamefont
  {Pehlevan}}]{canatar2021spectral}%
  \BibitemOpen
  \bibfield  {author} {\bibinfo {author} {\bibnamefont {Canatar}, \bibfnamefont
  {Abdulkadir}}, \bibinfo {author} {\bibfnamefont {Blake}\ \bibnamefont
  {Bordelon}}, and\ \bibinfo {author} {\bibfnamefont {Cengiz}\ \bibnamefont
  {Pehlevan}}} (\bibinfo {year} {2021}),\ \bibfield  {title} {\enquote
  {\bibinfo {title} {Spectral bias and task-model alignment explain
  generalization in kernel regression and infinitely wide neural networks},}\
  }\href@noop {} {\bibfield  {journal} {\bibinfo  {journal} {Nature
  communications}\ }\textbf {\bibinfo {volume} {12}}~(\bibinfo {number} {1}),\
  \bibinfo {pages} {2914}}\BibitemShut {NoStop}%
\bibitem [{\citenamefont {Canatar}\ \emph {et~al.}(2024)\citenamefont
  {Canatar}, \citenamefont {Feather}, \citenamefont {Wakhloo},\ and\
  \citenamefont {Chung}}]{canatar2024spectral}%
  \BibitemOpen
  \bibfield  {author} {\bibinfo {author} {\bibnamefont {Canatar}, \bibfnamefont
  {Abdulkadir}}, \bibinfo {author} {\bibfnamefont {Jenelle}\ \bibnamefont
  {Feather}}, \bibinfo {author} {\bibfnamefont {Albert}\ \bibnamefont
  {Wakhloo}}, and\ \bibinfo {author} {\bibfnamefont {SueYeon}\ \bibnamefont
  {Chung}}} (\bibinfo {year} {2024}),\ \bibfield  {title} {\enquote {\bibinfo
  {title} {A spectral theory of neural prediction and alignment},}\ }\href@noop
  {} {\bibfield  {journal} {\bibinfo  {journal} {Advances in Neural Information
  Processing Systems}\ }\textbf {\bibinfo {volume} {36}}}\BibitemShut {NoStop}%
\bibitem [{\citenamefont {Chklovskii}\ \emph {et~al.}(2002)\citenamefont
  {Chklovskii}, \citenamefont {Schikorski},\ and\ \citenamefont
  {Stevens}}]{chklovskii2002wiring}%
  \BibitemOpen
  \bibfield  {author} {\bibinfo {author} {\bibnamefont {Chklovskii},
  \bibfnamefont {Dmitri~B}}, \bibinfo {author} {\bibfnamefont {Thomas}\
  \bibnamefont {Schikorski}}, and\ \bibinfo {author} {\bibfnamefont
  {Charles~F.}\ \bibnamefont {Stevens}}} (\bibinfo {year} {2002}),\ \bibfield
  {title} {\enquote {\bibinfo {title} {Wiring optimization in cortical
  circuits},}\ }\href {https://doi.org/10.1016/S0896-6273(02)00679-7}
  {\bibfield  {journal} {\bibinfo  {journal} {Neuron}\ }\textbf {\bibinfo
  {volume} {34}}~(\bibinfo {number} {3}),\ \bibinfo {pages}
  {341--347}}\BibitemShut {NoStop}%
\bibitem [{\citenamefont {Chung}\ \emph {et~al.}(2018)\citenamefont {Chung},
  \citenamefont {Lee},\ and\ \citenamefont
  {Sompolinsky}}]{chung2018classification}%
  \BibitemOpen
  \bibfield  {author} {\bibinfo {author} {\bibnamefont {Chung}, \bibfnamefont
  {SueYeon}}, \bibinfo {author} {\bibfnamefont {Daniel~D.}\ \bibnamefont
  {Lee}}, and\ \bibinfo {author} {\bibfnamefont {Haim}\ \bibnamefont
  {Sompolinsky}}} (\bibinfo {year} {2018}),\ \bibfield  {title} {\enquote
  {\bibinfo {title} {Classification and geometry of general perceptual
  manifolds},}\ }\href {https://doi.org/10.1103/PhysRevX.8.031003} {\bibfield
  {journal} {\bibinfo  {journal} {Phys. Rev. X}\ }\textbf {\bibinfo {volume}
  {8}},\ \bibinfo {pages} {031003}}\BibitemShut {NoStop}%
\bibitem [{\citenamefont {Cohen}\ \emph {et~al.}(2020)\citenamefont {Cohen},
  \citenamefont {Chung}, \citenamefont {Lee},\ and\ \citenamefont
  {Sompolinsky}}]{cohen2020separability}%
  \BibitemOpen
  \bibfield  {author} {\bibinfo {author} {\bibnamefont {Cohen}, \bibfnamefont
  {Uri}}, \bibinfo {author} {\bibfnamefont {SueYeon}\ \bibnamefont {Chung}},
  \bibinfo {author} {\bibfnamefont {Daniel~D.}\ \bibnamefont {Lee}}, and\
  \bibinfo {author} {\bibfnamefont {Haim}\ \bibnamefont {Sompolinsky}}}
  (\bibinfo {year} {2020}),\ \bibfield  {title} {\enquote {\bibinfo {title}
  {Separability and geometry of object manifolds in deep neural networks},}\
  }\href {https://doi.org/10.1038/s41467-020-14578-5} {\bibfield  {journal}
  {\bibinfo  {journal} {Nature Communications}\ }\textbf {\bibinfo {volume}
  {11}}~(\bibinfo {number} {1}),\ \bibinfo {pages} {746}}\BibitemShut {NoStop}%
\bibitem [{\citenamefont {Cui}\ \emph {et~al.}(2023)\citenamefont {Cui},
  \citenamefont {Krzakala},\ and\ \citenamefont {Zdeborova}}]{cui2023bayes}%
  \BibitemOpen
  \bibfield  {author} {\bibinfo {author} {\bibnamefont {Cui}, \bibfnamefont
  {Hugo}}, \bibinfo {author} {\bibfnamefont {Florent}\ \bibnamefont
  {Krzakala}}, and\ \bibinfo {author} {\bibfnamefont {Lenka}\ \bibnamefont
  {Zdeborova}}} (\bibinfo {year} {2023}),\ \bibfield  {title} {\enquote
  {\bibinfo {title} {{B}ayes-optimal learning of deep random networks of
  extensive-width},}\ }in\ \href
  {https://proceedings.mlr.press/v202/cui23b.html} {\emph {\bibinfo {booktitle}
  {Proceedings of the 40th International Conference on Machine Learning}}},\
  \bibinfo {series} {Proceedings of Machine Learning Research}, Vol.\ \bibinfo
  {volume} {202},\ \bibinfo {editor} {edited by\ \bibinfo {editor}
  {\bibfnamefont {Andreas}\ \bibnamefont {Krause}}, \bibinfo {editor}
  {\bibfnamefont {Emma}\ \bibnamefont {Brunskill}}, \bibinfo {editor}
  {\bibfnamefont {Kyunghyun}\ \bibnamefont {Cho}}, \bibinfo {editor}
  {\bibfnamefont {Barbara}\ \bibnamefont {Engelhardt}}, \bibinfo {editor}
  {\bibfnamefont {Sivan}\ \bibnamefont {Sabato}}, \ and\ \bibinfo {editor}
  {\bibfnamefont {Jonathan}\ \bibnamefont {Scarlett}}}\ (\bibinfo  {publisher}
  {PMLR})\ pp.\ \bibinfo {pages} {6468--6521}\BibitemShut {NoStop}%
\bibitem [{\citenamefont {Engel}\ and\ \citenamefont {van~den
  Broeck}(2001)}]{engel2001statistical}%
  \BibitemOpen
  \bibfield  {author} {\bibinfo {author} {\bibnamefont {Engel}, \bibfnamefont
  {Andreas}}, and\ \bibinfo {author} {\bibfnamefont {Christian}\ \bibnamefont
  {van~den Broeck}}} (\bibinfo {year} {2001}),\ \href
  {https://doi.org/https://doi.org/10.1017/CBO9781139164542} {\emph {\bibinfo
  {title} {Statistical Mechanics of Learning}}}\ (\bibinfo  {publisher}
  {Cambridge University Press})\BibitemShut {NoStop}%
\bibitem [{\citenamefont {Farrell}\ \emph {et~al.}(2022)\citenamefont
  {Farrell}, \citenamefont {Bordelon}, \citenamefont {Trivedi},\ and\
  \citenamefont {Pehlevan}}]{farrell2022capacity}%
  \BibitemOpen
  \bibfield  {author} {\bibinfo {author} {\bibnamefont {Farrell}, \bibfnamefont
  {Matthew}}, \bibinfo {author} {\bibfnamefont {Blake}\ \bibnamefont
  {Bordelon}}, \bibinfo {author} {\bibfnamefont {Shubhendu}\ \bibnamefont
  {Trivedi}}, and\ \bibinfo {author} {\bibfnamefont {Cengiz}\ \bibnamefont
  {Pehlevan}}} (\bibinfo {year} {2022}),\ \bibfield  {title} {\enquote
  {\bibinfo {title} {Capacity of group-invariant linear readouts from
  equivariant representations: How many objects can be linearly classified
  under all possible views?}}\ }in\ \href
  {https://openreview.net/forum?id=_4GFbtOuWq-} {\emph {\bibinfo {booktitle}
  {International Conference on Learning Representations}}}\BibitemShut
  {NoStop}%
\bibitem [{\citenamefont {Fink}\ \emph {et~al.}(2025)\citenamefont {Fink},
  \citenamefont {Muscinelli}, \citenamefont {Wang}, \citenamefont {Hogan},
  \citenamefont {English}, \citenamefont {Axel}, \citenamefont {Litwin-Kumar},\
  and\ \citenamefont {Schoonover}}]{fink2025experience}%
  \BibitemOpen
  \bibfield  {author} {\bibinfo {author} {\bibnamefont {Fink}, \bibfnamefont
  {Andrew~JP}}, \bibinfo {author} {\bibfnamefont {Samuel~P.}\ \bibnamefont
  {Muscinelli}}, \bibinfo {author} {\bibfnamefont {Shuqi}\ \bibnamefont
  {Wang}}, \bibinfo {author} {\bibfnamefont {Marcus~I.}\ \bibnamefont {Hogan}},
  \bibinfo {author} {\bibfnamefont {Daniel~F.}\ \bibnamefont {English}},
  \bibinfo {author} {\bibfnamefont {Richard}\ \bibnamefont {Axel}}, \bibinfo
  {author} {\bibfnamefont {Ashok}\ \bibnamefont {Litwin-Kumar}}, and\ \bibinfo
  {author} {\bibfnamefont {Carl~E.}\ \bibnamefont {Schoonover}}} (\bibinfo
  {year} {2025}),\ \bibfield  {title} {\enquote {\bibinfo {title}
  {Experience-dependent reorganization of inhibitory neuron synaptic
  connectivity},}\ }\href {https://doi.org/10.1101/2025.01.16.633450}
  {\bibfield  {journal} {\bibinfo  {journal} {bioRxiv}\
  }10.1101/2025.01.16.633450},\ \Eprint
  {https://arxiv.org/abs/https://www.biorxiv.org/content/early/2025/01/16/2025.01.16.633450.full.pdf}
  {https://www.biorxiv.org/content/early/2025/01/16/2025.01.16.633450.full.pdf}
  \BibitemShut {NoStop}%
\bibitem [{\citenamefont {Gao}\ \emph {et~al.}(2017)\citenamefont {Gao},
  \citenamefont {Trautmann}, \citenamefont {Yu}, \citenamefont {Santhanam},
  \citenamefont {Ryu}, \citenamefont {Shenoy},\ and\ \citenamefont
  {Ganguli}}]{gao2017theory}%
  \BibitemOpen
  \bibfield  {author} {\bibinfo {author} {\bibnamefont {Gao}, \bibfnamefont
  {Peiran}}, \bibinfo {author} {\bibfnamefont {Eric}\ \bibnamefont
  {Trautmann}}, \bibinfo {author} {\bibfnamefont {Byron}\ \bibnamefont {Yu}},
  \bibinfo {author} {\bibfnamefont {Gopal}\ \bibnamefont {Santhanam}}, \bibinfo
  {author} {\bibfnamefont {Stephen}\ \bibnamefont {Ryu}}, \bibinfo {author}
  {\bibfnamefont {Krishna}\ \bibnamefont {Shenoy}}, and\ \bibinfo {author}
  {\bibfnamefont {Surya}\ \bibnamefont {Ganguli}}} (\bibinfo {year} {2017}),\
  \bibfield  {title} {\enquote {\bibinfo {title} {A theory of multineuronal
  dimensionality, dynamics and measurement},}\ }\href
  {https://doi.org/10.1101/214262} {\bibfield  {journal} {\bibinfo  {journal}
  {bioRxiv}\ }10.1101/214262},\ \Eprint
  {https://arxiv.org/abs/https://www.biorxiv.org/content/early/2017/11/12/214262.full.pdf}
  {https://www.biorxiv.org/content/early/2017/11/12/214262.full.pdf}
  \BibitemShut {NoStop}%
\bibitem [{\citenamefont {Goldt}\ \emph {et~al.}(2019)\citenamefont {Goldt},
  \citenamefont {Advani}, \citenamefont {Saxe}, \citenamefont {Krzakala},\ and\
  \citenamefont {Zdeborov{\'a}}}]{goldt2019dynamics}%
  \BibitemOpen
  \bibfield  {author} {\bibinfo {author} {\bibnamefont {Goldt}, \bibfnamefont
  {Sebastian}}, \bibinfo {author} {\bibfnamefont {Madhu}\ \bibnamefont
  {Advani}}, \bibinfo {author} {\bibfnamefont {Andrew~M}\ \bibnamefont {Saxe}},
  \bibinfo {author} {\bibfnamefont {Florent}\ \bibnamefont {Krzakala}}, and\
  \bibinfo {author} {\bibfnamefont {Lenka}\ \bibnamefont {Zdeborov{\'a}}}}
  (\bibinfo {year} {2019}),\ \bibfield  {title} {\enquote {\bibinfo {title}
  {Dynamics of stochastic gradient descent for two-layer neural networks in the
  teacher-student setup},}\ }\href@noop {} {\bibfield  {journal} {\bibinfo
  {journal} {Advances in neural information processing systems}\ }\textbf
  {\bibinfo {volume} {32}}}\BibitemShut {NoStop}%
\bibitem [{\citenamefont {Goldt}\ \emph {et~al.}(2020)\citenamefont {Goldt},
  \citenamefont {M\'ezard}, \citenamefont {Krzakala},\ and\ \citenamefont
  {Zdeborov\'a}}]{goldt2020hidden}%
  \BibitemOpen
  \bibfield  {author} {\bibinfo {author} {\bibnamefont {Goldt}, \bibfnamefont
  {Sebastian}}, \bibinfo {author} {\bibfnamefont {Marc}\ \bibnamefont
  {M\'ezard}}, \bibinfo {author} {\bibfnamefont {Florent}\ \bibnamefont
  {Krzakala}}, and\ \bibinfo {author} {\bibfnamefont {Lenka}\ \bibnamefont
  {Zdeborov\'a}}} (\bibinfo {year} {2020}),\ \bibfield  {title} {\enquote
  {\bibinfo {title} {Modeling the influence of data structure on learning in
  neural networks: The hidden manifold model},}\ }\href
  {https://doi.org/10.1103/PhysRevX.10.041044} {\bibfield  {journal} {\bibinfo
  {journal} {Physical Review X}\ }\textbf {\bibinfo {volume} {10}},\ \bibinfo
  {pages} {041044}}\BibitemShut {NoStop}%
\bibitem [{\citenamefont {Golikov}\ and\ \citenamefont
  {Yang}(2022)}]{golikov2022nongaussian}%
  \BibitemOpen
  \bibfield  {author} {\bibinfo {author} {\bibnamefont {Golikov}, \bibfnamefont
  {Eugene}}, and\ \bibinfo {author} {\bibfnamefont {Greg}\ \bibnamefont
  {Yang}}} (\bibinfo {year} {2022}),\ \bibfield  {title} {\enquote {\bibinfo
  {title} {Non-{G}aussian {T}ensor {P}rograms},}\ }in\ \href
  {https://proceedings.neurips.cc/paper_files/paper/2022/file/8707924df5e207fa496f729f49069446-Paper-Conference.pdf}
  {\emph {\bibinfo {booktitle} {Advances in Neural Information Processing
  Systems}}},\ Vol.~\bibinfo {volume} {35},\ \bibinfo {editor} {edited by\
  \bibinfo {editor} {\bibfnamefont {S.}~\bibnamefont {Koyejo}}, \bibinfo
  {editor} {\bibfnamefont {S.}~\bibnamefont {Mohamed}}, \bibinfo {editor}
  {\bibfnamefont {A.}~\bibnamefont {Agarwal}}, \bibinfo {editor} {\bibfnamefont
  {D.}~\bibnamefont {Belgrave}}, \bibinfo {editor} {\bibfnamefont
  {K.}~\bibnamefont {Cho}}, \ and\ \bibinfo {editor} {\bibfnamefont
  {A.}~\bibnamefont {Oh}}}\ (\bibinfo  {publisher} {Curran Associates, Inc.})\
  pp.\ \bibinfo {pages} {21521--21533}\BibitemShut {NoStop}%
\bibitem [{\citenamefont {Hara}\ \emph {et~al.}(2011)\citenamefont {Hara},
  \citenamefont {Katahira}, \citenamefont {Okanoya},\ and\ \citenamefont
  {Okada}}]{hara2011node}%
  \BibitemOpen
  \bibfield  {author} {\bibinfo {author} {\bibnamefont {Hara}, \bibfnamefont
  {Kazuyuki}}, \bibinfo {author} {\bibfnamefont {Kentaro}\ \bibnamefont
  {Katahira}}, \bibinfo {author} {\bibfnamefont {Kazuo}\ \bibnamefont
  {Okanoya}}, and\ \bibinfo {author} {\bibfnamefont {Masato}\ \bibnamefont
  {Okada}}} (\bibinfo {year} {2011}),\ \bibfield  {title} {\enquote {\bibinfo
  {title} {Statistical mechanics of on-line node-perturbation learning},}\
  }\href {https://doi.org/10.2197/ipsjtrans.4.23} {\bibfield  {journal}
  {\bibinfo  {journal} {IPSJ Online Transactions}\ }\textbf {\bibinfo {volume}
  {4}},\ \bibinfo {pages} {23--32}}\BibitemShut {NoStop}%
\bibitem [{\citenamefont {Hara}\ \emph {et~al.}(2013)\citenamefont {Hara},
  \citenamefont {Katahira}, \citenamefont {Okanoya},\ and\ \citenamefont
  {Okada}}]{hara2013node}%
  \BibitemOpen
  \bibfield  {author} {\bibinfo {author} {\bibnamefont {Hara}, \bibfnamefont
  {Kazuyuki}}, \bibinfo {author} {\bibfnamefont {Kentaro}\ \bibnamefont
  {Katahira}}, \bibinfo {author} {\bibfnamefont {Kazuo}\ \bibnamefont
  {Okanoya}}, and\ \bibinfo {author} {\bibfnamefont {Masato}\ \bibnamefont
  {Okada}}} (\bibinfo {year} {2013}),\ \bibfield  {title} {\enquote {\bibinfo
  {title} {Statistical mechanics of node-perturbation learning for nonlinear
  perceptron},}\ }\href {https://doi.org/10.7566/JPSJ.82.054001} {\bibfield
  {journal} {\bibinfo  {journal} {Journal of the Physical Society of Japan}\
  }\textbf {\bibinfo {volume} {82}}~(\bibinfo {number} {5}),\ \bibinfo {pages}
  {054001}}\BibitemShut {NoStop}%
\bibitem [{\citenamefont {Harvey}\ \emph {et~al.}(2024)\citenamefont {Harvey},
  \citenamefont {Lipshutz},\ and\ \citenamefont {Williams}}]{harvey2024what}%
  \BibitemOpen
  \bibfield  {author} {\bibinfo {author} {\bibnamefont {Harvey}, \bibfnamefont
  {Sarah~E}}, \bibinfo {author} {\bibfnamefont {David}\ \bibnamefont
  {Lipshutz}}, and\ \bibinfo {author} {\bibfnamefont {Alex~H}\ \bibnamefont
  {Williams}}} (\bibinfo {year} {2024}),\ \bibfield  {title} {\enquote
  {\bibinfo {title} {What representational similarity measures imply about
  decodable information},}\ }in\ \href
  {https://openreview.net/forum?id=hqfzH6GCYj} {\emph {\bibinfo {booktitle}
  {UniReps: 2nd Edition of the Workshop on Unifying Representations in Neural
  Models}}}\BibitemShut {NoStop}%
\bibitem [{\citenamefont {Hastie}\ \emph {et~al.}(2022)\citenamefont {Hastie},
  \citenamefont {Montanari}, \citenamefont {Rosset},\ and\ \citenamefont
  {Tibshirani}}]{hastie2022surprises}%
  \BibitemOpen
  \bibfield  {author} {\bibinfo {author} {\bibnamefont {Hastie}, \bibfnamefont
  {Trevor}}, \bibinfo {author} {\bibfnamefont {Andrea}\ \bibnamefont
  {Montanari}}, \bibinfo {author} {\bibfnamefont {Saharon}\ \bibnamefont
  {Rosset}}, and\ \bibinfo {author} {\bibfnamefont {Ryan~J}\ \bibnamefont
  {Tibshirani}}} (\bibinfo {year} {2022}),\ \bibfield  {title} {\enquote
  {\bibinfo {title} {Surprises in high-dimensional ridgeless least squares
  interpolation},}\ }\href@noop {} {\bibfield  {journal} {\bibinfo  {journal}
  {The Annals of Statistics}\ }\textbf {\bibinfo {volume} {50}}~(\bibinfo
  {number} {2}),\ \bibinfo {pages} {949--986}}\BibitemShut {NoStop}%
\bibitem [{\citenamefont {Hebb}(2005)}]{hebb2005organization}%
  \BibitemOpen
  \bibfield  {author} {\bibinfo {author} {\bibnamefont {Hebb}, \bibfnamefont
  {Donald~Olding}}} (\bibinfo {year} {2005}),\ \href@noop {} {\emph {\bibinfo
  {title} {The organization of behavior: A neuropsychological theory}}}\
  (\bibinfo  {publisher} {Psychology press})\BibitemShut {NoStop}%
\bibitem [{\citenamefont {Hirokawa}\ \emph {et~al.}(2019)\citenamefont
  {Hirokawa}, \citenamefont {Vaughan}, \citenamefont {Masset}, \citenamefont
  {Ott},\ and\ \citenamefont {Kepecs}}]{hirokawa2019frontal}%
  \BibitemOpen
  \bibfield  {author} {\bibinfo {author} {\bibnamefont {Hirokawa},
  \bibfnamefont {Junya}}, \bibinfo {author} {\bibfnamefont {Alexander}\
  \bibnamefont {Vaughan}}, \bibinfo {author} {\bibfnamefont {Paul}\
  \bibnamefont {Masset}}, \bibinfo {author} {\bibfnamefont {Torben}\
  \bibnamefont {Ott}}, and\ \bibinfo {author} {\bibfnamefont {Adam}\
  \bibnamefont {Kepecs}}} (\bibinfo {year} {2019}),\ \bibfield  {title}
  {\enquote {\bibinfo {title} {Frontal cortex neuron types categorically encode
  single decision variables},}\ }\href
  {https://doi.org/10.1038/s41586-019-1816-9} {\bibfield  {journal} {\bibinfo
  {journal} {Nature}\ }\textbf {\bibinfo {volume} {576}}~(\bibinfo {number}
  {7787}),\ \bibinfo {pages} {446--451}}\BibitemShut {NoStop}%
\bibitem [{\citenamefont {Hu}\ and\ \citenamefont
  {Lu}(2022)}]{hu2022universality}%
  \BibitemOpen
  \bibfield  {author} {\bibinfo {author} {\bibnamefont {Hu}, \bibfnamefont
  {Hong}}, and\ \bibinfo {author} {\bibfnamefont {Yue~M}\ \bibnamefont {Lu}}}
  (\bibinfo {year} {2022}),\ \bibfield  {title} {\enquote {\bibinfo {title}
  {Universality laws for high-dimensional learning with random features},}\
  }\href@noop {} {\bibfield  {journal} {\bibinfo  {journal} {IEEE Transactions
  on Information Theory}\ }\textbf {\bibinfo {volume} {69}}~(\bibinfo {number}
  {3}),\ \bibinfo {pages} {1932--1964}}\BibitemShut {NoStop}%
\bibitem [{\citenamefont {Jacot}\ \emph {et~al.}(2018)\citenamefont {Jacot},
  \citenamefont {Gabriel},\ and\ \citenamefont {Hongler}}]{jacot2018neural}%
  \BibitemOpen
  \bibfield  {author} {\bibinfo {author} {\bibnamefont {Jacot}, \bibfnamefont
  {Arthur}}, \bibinfo {author} {\bibfnamefont {Franck}\ \bibnamefont
  {Gabriel}}, and\ \bibinfo {author} {\bibfnamefont {Cl{\'e}ment}\ \bibnamefont
  {Hongler}}} (\bibinfo {year} {2018}),\ \bibfield  {title} {\enquote {\bibinfo
  {title} {Neural tangent kernel: Convergence and generalization in neural
  networks},}\ }\href@noop {} {\bibfield  {journal} {\bibinfo  {journal}
  {Advances in neural information processing systems}\ }\textbf {\bibinfo
  {volume} {31}}}\BibitemShut {NoStop}%
\bibitem [{\citenamefont {Kandler}\ \emph {et~al.}(2009)\citenamefont
  {Kandler}, \citenamefont {Clause},\ and\ \citenamefont
  {Noh}}]{kandler2009tonotopic}%
  \BibitemOpen
  \bibfield  {author} {\bibinfo {author} {\bibnamefont {Kandler}, \bibfnamefont
  {Karl}}, \bibinfo {author} {\bibfnamefont {Amanda}\ \bibnamefont {Clause}},
  and\ \bibinfo {author} {\bibfnamefont {Jihyun}\ \bibnamefont {Noh}}}
  (\bibinfo {year} {2009}),\ \bibfield  {title} {\enquote {\bibinfo {title}
  {Tonotopic reorganization of developing auditory brainstem circuits},}\
  }\href {https://doi.org/https://doi.org/10.1038/nn.2332} {\bibfield
  {journal} {\bibinfo  {journal} {Nature Neuroscience}\ }\textbf {\bibinfo
  {volume} {12}}~(\bibinfo {number} {6}),\ \bibinfo {pages}
  {711--717}}\BibitemShut {NoStop}%
\bibitem [{\citenamefont {Kang}\ \emph {et~al.}(2025)\citenamefont {Kang},
  \citenamefont {Canatar},\ and\ \citenamefont {Chung}}]{kang2025spectral}%
  \BibitemOpen
  \bibfield  {author} {\bibinfo {author} {\bibnamefont {Kang}, \bibfnamefont
  {Hyunmo}}, \bibinfo {author} {\bibfnamefont {Abdulkadir}\ \bibnamefont
  {Canatar}}, and\ \bibinfo {author} {\bibfnamefont {SueYeon}\ \bibnamefont
  {Chung}}} (\bibinfo {year} {2025}),\ \bibfield  {title} {\enquote {\bibinfo
  {title} {Spectral analysis of representational similarity with limited
  neurons},}\ }\href {https://arxiv.org/abs/2502.19648} {\bibfield  {journal}
  {\bibinfo  {journal} {arXiv}\ }}\Eprint {https://arxiv.org/abs/2502.19648}
  {arXiv:2502.19648 [cond-mat.dis-nn]} \BibitemShut {NoStop}%
\bibitem [{\citenamefont {Khaligh-Razavi}\ and\ \citenamefont
  {Kriegeskorte}(2014)}]{kaligh2014cortical}%
  \BibitemOpen
  \bibfield  {author} {\bibinfo {author} {\bibnamefont {Khaligh-Razavi},
  \bibfnamefont {Seyed-Mahdi}}, and\ \bibinfo {author} {\bibfnamefont
  {Nikolaus}\ \bibnamefont {Kriegeskorte}}} (\bibinfo {year} {2014}),\
  \bibfield  {title} {\enquote {\bibinfo {title} {Deep supervised, but not
  unsupervised, models may explain it cortical representation},}\ }\href
  {https://doi.org/10.1371/journal.pcbi.1003915} {\bibfield  {journal}
  {\bibinfo  {journal} {PLOS Computational Biology}\ }\textbf {\bibinfo
  {volume} {10}},\ \bibinfo {pages} {1--29}}\BibitemShut {NoStop}%
\bibitem [{\citenamefont {Khona}\ \emph {et~al.}(2025)\citenamefont {Khona},
  \citenamefont {Chandra},\ and\ \citenamefont {Fiete}}]{khona2025modules}%
  \BibitemOpen
  \bibfield  {author} {\bibinfo {author} {\bibnamefont {Khona}, \bibfnamefont
  {Mikail}}, \bibinfo {author} {\bibfnamefont {Sarthak}\ \bibnamefont
  {Chandra}}, and\ \bibinfo {author} {\bibfnamefont {Ila}\ \bibnamefont
  {Fiete}}} (\bibinfo {year} {2025}),\ \bibfield  {title} {\enquote {\bibinfo
  {title} {Global modules robustly emerge from local interactions and smooth
  gradients},}\ }\href {https://doi.org/10.1038/s41586-024-08541-3} {\bibfield
  {journal} {\bibinfo  {journal} {Nature}\
  }10.1038/s41586-024-08541-3}\BibitemShut {NoStop}%
\bibitem [{\citenamefont {Krakauer}\ \emph {et~al.}(2017)\citenamefont
  {Krakauer}, \citenamefont {Ghazanfar}, \citenamefont {Gomez-Marin},
  \citenamefont {MacIver},\ and\ \citenamefont
  {Poeppel}}]{krakauer2017behavior}%
  \BibitemOpen
  \bibfield  {author} {\bibinfo {author} {\bibnamefont {Krakauer},
  \bibfnamefont {John~W}}, \bibinfo {author} {\bibfnamefont {Asif~A.}\
  \bibnamefont {Ghazanfar}}, \bibinfo {author} {\bibfnamefont {Alex}\
  \bibnamefont {Gomez-Marin}}, \bibinfo {author} {\bibfnamefont {Malcolm~A.}\
  \bibnamefont {MacIver}}, and\ \bibinfo {author} {\bibfnamefont {David}\
  \bibnamefont {Poeppel}}} (\bibinfo {year} {2017}),\ \bibfield  {title}
  {\enquote {\bibinfo {title} {Neuroscience needs behavior: Correcting a
  reductionist bias},}\ }\href
  {https://doi.org/https://doi.org/10.1016/j.neuron.2016.12.041} {\bibfield
  {journal} {\bibinfo  {journal} {Neuron}\ }\textbf {\bibinfo {volume}
  {93}}~(\bibinfo {number} {3}),\ \bibinfo {pages} {480--490}}\BibitemShut
  {NoStop}%
\bibitem [{\citenamefont {Kriegeskorte}\ \emph {et~al.}(2008)\citenamefont
  {Kriegeskorte}, \citenamefont {Mur},\ and\ \citenamefont
  {Bandettini}}]{kriegeskorte2008rsa}%
  \BibitemOpen
  \bibfield  {author} {\bibinfo {author} {\bibnamefont {Kriegeskorte},
  \bibfnamefont {Nikolaus}}, \bibinfo {author} {\bibfnamefont {Marieke}\
  \bibnamefont {Mur}}, and\ \bibinfo {author} {\bibfnamefont {Peter~A.}\
  \bibnamefont {Bandettini}}} (\bibinfo {year} {2008}),\ \bibfield  {title}
  {\enquote {\bibinfo {title} {Representational similarity analysis -
  connecting the branches of systems neuroscience},}\ }\href
  {https://doi.org/10.3389/neuro.06.004.2008} {\bibfield  {journal} {\bibinfo
  {journal} {Frontiers in Systems Neuroscience}\ }\textbf {\bibinfo {volume}
  {Volume 2 - 2008}},\ 10.3389/neuro.06.004.2008}\BibitemShut {NoStop}%
\bibitem [{\citenamefont {Kriegeskorte}\ and\ \citenamefont
  {Wei}(2021)}]{kriegeskorte2021geometry}%
  \BibitemOpen
  \bibfield  {author} {\bibinfo {author} {\bibnamefont {Kriegeskorte},
  \bibfnamefont {Nikolaus}}, and\ \bibinfo {author} {\bibfnamefont {Xue-Xin}\
  \bibnamefont {Wei}}} (\bibinfo {year} {2021}),\ \bibfield  {title} {\enquote
  {\bibinfo {title} {Neural tuning and representational geometry},}\ }\href
  {https://doi.org/10.1038/s41583-021-00502-3} {\bibfield  {journal} {\bibinfo
  {journal} {Nature Reviews Neuroscience}\ }\textbf {\bibinfo {volume}
  {22}}~(\bibinfo {number} {11}),\ \bibinfo {pages} {703--718}}\BibitemShut
  {NoStop}%
\bibitem [{\citenamefont {Lee}\ \emph {et~al.}(2019)\citenamefont {Lee},
  \citenamefont {Xiao}, \citenamefont {Schoenholz}, \citenamefont {Bahri},
  \citenamefont {Novak}, \citenamefont {Sohl-Dickstein},\ and\ \citenamefont
  {Pennington}}]{lee2019wide}%
  \BibitemOpen
  \bibfield  {author} {\bibinfo {author} {\bibnamefont {Lee}, \bibfnamefont
  {Jaehoon}}, \bibinfo {author} {\bibfnamefont {Lechao}\ \bibnamefont {Xiao}},
  \bibinfo {author} {\bibfnamefont {Samuel}\ \bibnamefont {Schoenholz}},
  \bibinfo {author} {\bibfnamefont {Yasaman}\ \bibnamefont {Bahri}}, \bibinfo
  {author} {\bibfnamefont {Roman}\ \bibnamefont {Novak}}, \bibinfo {author}
  {\bibfnamefont {Jascha}\ \bibnamefont {Sohl-Dickstein}}, and\ \bibinfo
  {author} {\bibfnamefont {Jeffrey}\ \bibnamefont {Pennington}}} (\bibinfo
  {year} {2019}),\ \bibfield  {title} {\enquote {\bibinfo {title} {Wide neural
  networks of any depth evolve as linear models under gradient descent},}\ }in\
  \href
  {https://proceedings.neurips.cc/paper_files/paper/2019/file/0d1a9651497a38d8b1c3871c84528bd4-Paper.pdf}
  {\emph {\bibinfo {booktitle} {Advances in Neural Information Processing
  Systems}}},\ Vol.~\bibinfo {volume} {32},\ \bibinfo {editor} {edited by\
  \bibinfo {editor} {\bibfnamefont {H.}~\bibnamefont {Wallach}}, \bibinfo
  {editor} {\bibfnamefont {H.}~\bibnamefont {Larochelle}}, \bibinfo {editor}
  {\bibfnamefont {A.}~\bibnamefont {Beygelzimer}}, \bibinfo {editor}
  {\bibfnamefont {F.}~\bibnamefont {d\textquotesingle Alch\'{e}-Buc}}, \bibinfo
  {editor} {\bibfnamefont {E.}~\bibnamefont {Fox}}, \ and\ \bibinfo {editor}
  {\bibfnamefont {R.}~\bibnamefont {Garnett}}}\ (\bibinfo  {publisher} {Curran
  Associates, Inc.})\BibitemShut {NoStop}%
\bibitem [{\citenamefont {Lillicrap}\ \emph {et~al.}(2016)\citenamefont
  {Lillicrap}, \citenamefont {Cownden}, \citenamefont {Tweed},\ and\
  \citenamefont {Akerman}}]{lillicrap2016random}%
  \BibitemOpen
  \bibfield  {author} {\bibinfo {author} {\bibnamefont {Lillicrap},
  \bibfnamefont {Timothy~P}}, \bibinfo {author} {\bibfnamefont {Daniel}\
  \bibnamefont {Cownden}}, \bibinfo {author} {\bibfnamefont {Douglas~B}\
  \bibnamefont {Tweed}}, and\ \bibinfo {author} {\bibfnamefont {Colin~J}\
  \bibnamefont {Akerman}}} (\bibinfo {year} {2016}),\ \bibfield  {title}
  {\enquote {\bibinfo {title} {Random synaptic feedback weights support error
  backpropagation for deep learning},}\ }\href@noop {} {\bibfield  {journal}
  {\bibinfo  {journal} {Nature communications}\ }\textbf {\bibinfo {volume}
  {7}}~(\bibinfo {number} {1}),\ \bibinfo {pages} {13276}}\BibitemShut
  {NoStop}%
\bibitem [{\citenamefont {Marchenko}\ and\ \citenamefont
  {Pastur}(1967)}]{marchenko1967distribution}%
  \BibitemOpen
  \bibfield  {author} {\bibinfo {author} {\bibnamefont {Marchenko},
  \bibfnamefont {Vladimir~Alexandrovich}}, and\ \bibinfo {author}
  {\bibfnamefont {Leonid~Andreevich}\ \bibnamefont {Pastur}}} (\bibinfo {year}
  {1967}),\ \bibfield  {title} {\enquote {\bibinfo {title} {Distribution of
  eigenvalues for some sets of random matrices},}\ }\href@noop {} {\bibfield
  {journal} {\bibinfo  {journal} {Matematicheskii Sbornik}\ }\textbf {\bibinfo
  {volume} {114}}~(\bibinfo {number} {4}),\ \bibinfo {pages}
  {507--536}}\BibitemShut {NoStop}%
\bibitem [{\citenamefont {Masset}\ \emph {et~al.}(2022)\citenamefont {Masset},
  \citenamefont {Qin},\ and\ \citenamefont
  {Zavatone-Veth}}]{masset2022drifting}%
  \BibitemOpen
  \bibfield  {author} {\bibinfo {author} {\bibnamefont {Masset}, \bibfnamefont
  {Paul}}, \bibinfo {author} {\bibfnamefont {Shanshan}\ \bibnamefont {Qin}},
  and\ \bibinfo {author} {\bibfnamefont {Jacob~A}\ \bibnamefont
  {Zavatone-Veth}}} (\bibinfo {year} {2022}),\ \bibfield  {title} {\enquote
  {\bibinfo {title} {Drifting neuronal representations: Bug or feature?}}\
  }\href {https://doi.org/doi.org/10.1007/s00422-021-00916-3} {\bibinfo
  {journal} {Biological Cybernetics}\ ,\ \bibinfo {pages} {1--14}}\BibitemShut
  {NoStop}%
\bibitem [{\citenamefont {van Meegen}\ and\ \citenamefont
  {Sompolinsky}(2025)}]{vanmeegen2025coding}%
  \BibitemOpen
\bibfield  {journal} {  }\bibfield  {author} {\bibinfo {author} {\bibnamefont
  {van Meegen}, \bibfnamefont {Alexander}}, and\ \bibinfo {author}
  {\bibfnamefont {Haim}\ \bibnamefont {Sompolinsky}}} (\bibinfo {year}
  {2025}),\ \bibfield  {title} {\enquote {\bibinfo {title} {Coding schemes in
  neural networks learning classification tasks},}\ }\href
  {https://doi.org/10.1038/s41467-025-58276-6} {\bibfield  {journal} {\bibinfo
  {journal} {Nature Communications}\ }\textbf {\bibinfo {volume}
  {16}}~(\bibinfo {number} {1}),\ \bibinfo {pages} {3354}}\BibitemShut
  {NoStop}%
\bibitem [{\citenamefont {M{\'e}zard}\ \emph {et~al.}(1987)\citenamefont
  {M{\'e}zard}, \citenamefont {Parisi},\ and\ \citenamefont
  {Virasoro}}]{mezard1987spin}%
  \BibitemOpen
  \bibfield  {author} {\bibinfo {author} {\bibnamefont {M{\'e}zard},
  \bibfnamefont {Marc}}, \bibinfo {author} {\bibfnamefont {Giorgio}\
  \bibnamefont {Parisi}}, and\ \bibinfo {author} {\bibfnamefont {Miguel~Angel}\
  \bibnamefont {Virasoro}}} (\bibinfo {year} {1987}),\ \href
  {https://doi.org/https://doi.org/10.1142/0271} {\emph {\bibinfo {title} {Spin
  Glass Theory and Beyond: An Introduction to the Replica Method and Its
  Applications}}}\ (\bibinfo  {publisher} {World Scientific Publishing
  Company})\BibitemShut {NoStop}%
\bibitem [{\citenamefont {Mignacco}\ and\ \citenamefont
  {Mori}(2025)}]{mignacco2025optimal}%
  \BibitemOpen
  \bibfield  {author} {\bibinfo {author} {\bibnamefont {Mignacco},
  \bibfnamefont {Francesca}}, and\ \bibinfo {author} {\bibfnamefont
  {Francesco}\ \bibnamefont {Mori}}} (\bibinfo {year} {2025}),\ \bibfield
  {title} {\enquote {\bibinfo {title} {A statistical physics framework for
  optimal learning},}\ }\href {https://doi.org/10.48550/arXiv.2507.07907}
  {\bibfield  {journal} {\bibinfo  {journal} {arXiv}\
  }10.48550/arXiv.2507.07907},\ \Eprint {https://arxiv.org/abs/2507.07907}
  {arXiv:2507.07907 [cond-mat.dis-nn]} \BibitemShut {NoStop}%
\bibitem [{\citenamefont {Misiakiewicz}\ and\ \citenamefont
  {Saeed}(2024)}]{misiakiewicz2024non}%
  \BibitemOpen
  \bibfield  {author} {\bibinfo {author} {\bibnamefont {Misiakiewicz},
  \bibfnamefont {Theodor}}, and\ \bibinfo {author} {\bibfnamefont {Basil}\
  \bibnamefont {Saeed}}} (\bibinfo {year} {2024}),\ \bibfield  {title}
  {\enquote {\bibinfo {title} {A non-asymptotic theory of kernel ridge
  regression: deterministic equivalents, test error, and {GCV} estimator},}\
  }\href@noop {} {\bibinfo  {journal} {arXiv preprint arXiv:2403.08938}\
  }\BibitemShut {NoStop}%
\bibitem [{\citenamefont {Montanari}\ and\ \citenamefont
  {Urbani}(2025)}]{montanari2025dynamical}%
  \BibitemOpen
\bibfield  {journal} {  }\bibfield  {author} {\bibinfo {author} {\bibnamefont
  {Montanari}, \bibfnamefont {Andrea}}, and\ \bibinfo {author} {\bibfnamefont
  {Pierfrancesco}\ \bibnamefont {Urbani}}} (\bibinfo {year} {2025}),\ \bibfield
   {title} {\enquote {\bibinfo {title} {Dynamical decoupling of generalization
  and overfitting in large two-layer networks},}\ }\href@noop {} {\bibinfo
  {journal} {arXiv preprint arXiv:2502.21269}\ }\BibitemShut {NoStop}%
\bibitem [{\citenamefont {Mori}\ \emph {et~al.}(2025)\citenamefont {Mori},
  \citenamefont {Mannelli},\ and\ \citenamefont {Mignacco}}]{mori2025optimal}%
  \BibitemOpen
\bibfield  {journal} {  }\bibfield  {author} {\bibinfo {author} {\bibnamefont
  {Mori}, \bibfnamefont {Francesco}}, \bibinfo {author} {\bibfnamefont
  {Stefano~Sarao}\ \bibnamefont {Mannelli}}, and\ \bibinfo {author}
  {\bibfnamefont {Francesca}\ \bibnamefont {Mignacco}}} (\bibinfo {year}
  {2025}),\ \bibfield  {title} {\enquote {\bibinfo {title} {Optimal protocols
  for continual learning via statistical physics and control theory},}\ }in\
  \href {https://openreview.net/forum?id=rhhQjGj09A} {\emph {\bibinfo
  {booktitle} {The Thirteenth International Conference on Learning
  Representations}}}\BibitemShut {NoStop}%
\bibitem [{\citenamefont {Murthy}(2011)}]{murthy2011maps}%
  \BibitemOpen
  \bibfield  {author} {\bibinfo {author} {\bibnamefont {Murthy}, \bibfnamefont
  {Venkatesh~N}}} (\bibinfo {year} {2011}),\ \bibfield  {title} {\enquote
  {\bibinfo {title} {Olfactory maps in the brain},}\ }\href
  {https://doi.org/10.1146/annurev-neuro-061010-113738} {\bibfield  {journal}
  {\bibinfo  {journal} {Annual Review of Neuroscience}\ }\textbf {\bibinfo
  {volume} {34}}~(\bibinfo {number} {1}),\ \bibinfo {pages}
  {233--258}}\BibitemShut {NoStop}%
\bibitem [{\citenamefont {N{\o}kland}(2016)}]{nokland2016direct}%
  \BibitemOpen
  \bibfield  {author} {\bibinfo {author} {\bibnamefont {N{\o}kland},
  \bibfnamefont {Arild}}} (\bibinfo {year} {2016}),\ \bibfield  {title}
  {\enquote {\bibinfo {title} {Direct feedback alignment provides learning in
  deep neural networks},}\ }\href@noop {} {\bibfield  {journal} {\bibinfo
  {journal} {Advances in neural information processing systems}\ }\textbf
  {\bibinfo {volume} {29}}}\BibitemShut {NoStop}%
\bibitem [{\citenamefont {Pashakhanloo}\ and\ \citenamefont
  {Koulakov}(2023)}]{pashakhanloo2023drift}%
  \BibitemOpen
  \bibfield  {author} {\bibinfo {author} {\bibnamefont {Pashakhanloo},
  \bibfnamefont {Farhad}}, and\ \bibinfo {author} {\bibfnamefont {Alexei}\
  \bibnamefont {Koulakov}}} (\bibinfo {year} {2023}),\ \bibfield  {title}
  {\enquote {\bibinfo {title} {Stochastic gradient descent-induced drift of
  representation in a two-layer neural network},}\ }in\ \href
  {https://proceedings.mlr.press/v202/pashakhanloo23a.html} {\emph {\bibinfo
  {booktitle} {Proceedings of the 40th International Conference on Machine
  Learning}}},\ \bibinfo {series} {Proceedings of Machine Learning Research},
  Vol.\ \bibinfo {volume} {202},\ \bibinfo {editor} {edited by\ \bibinfo
  {editor} {\bibfnamefont {Andreas}\ \bibnamefont {Krause}}, \bibinfo {editor}
  {\bibfnamefont {Emma}\ \bibnamefont {Brunskill}}, \bibinfo {editor}
  {\bibfnamefont {Kyunghyun}\ \bibnamefont {Cho}}, \bibinfo {editor}
  {\bibfnamefont {Barbara}\ \bibnamefont {Engelhardt}}, \bibinfo {editor}
  {\bibfnamefont {Sivan}\ \bibnamefont {Sabato}}, \ and\ \bibinfo {editor}
  {\bibfnamefont {Jonathan}\ \bibnamefont {Scarlett}}}\ (\bibinfo  {publisher}
  {PMLR})\ pp.\ \bibinfo {pages} {27401--27419}\BibitemShut {NoStop}%
\bibitem [{\citenamefont {Qin}\ \emph {et~al.}(2023)\citenamefont {Qin},
  \citenamefont {Farashahi}, \citenamefont {Lipshutz}, \citenamefont
  {Sengupta}, \citenamefont {Chklovskii},\ and\ \citenamefont
  {Pehlevan}}]{qin2023coordinated}%
  \BibitemOpen
  \bibfield  {author} {\bibinfo {author} {\bibnamefont {Qin}, \bibfnamefont
  {Shanshan}}, \bibinfo {author} {\bibfnamefont {Shiva}\ \bibnamefont
  {Farashahi}}, \bibinfo {author} {\bibfnamefont {David}\ \bibnamefont
  {Lipshutz}}, \bibinfo {author} {\bibfnamefont {Anirvan~M.}\ \bibnamefont
  {Sengupta}}, \bibinfo {author} {\bibfnamefont {Dmitri~B.}\ \bibnamefont
  {Chklovskii}}, and\ \bibinfo {author} {\bibfnamefont {Cengiz}\ \bibnamefont
  {Pehlevan}}} (\bibinfo {year} {2023}),\ \bibfield  {title} {\enquote
  {\bibinfo {title} {Coordinated drift of receptive fields in
  {Hebbian/anti-Hebbian} network models during noisy representation
  learning},}\ }\href {https://doi.org/10.1038/s41593-022-01225-z} {\bibfield
  {journal} {\bibinfo  {journal} {Nature Neuroscience}\ }\textbf {\bibinfo
  {volume} {26}}~(\bibinfo {number} {2}),\ \bibinfo {pages}
  {339--349}}\BibitemShut {NoStop}%
\bibitem [{\citenamefont {Rule}\ \emph {et~al.}(2019)\citenamefont {Rule},
  \citenamefont {O’Leary},\ and\ \citenamefont {Harvey}}]{rule2019causes}%
  \BibitemOpen
  \bibfield  {author} {\bibinfo {author} {\bibnamefont {Rule}, \bibfnamefont
  {Michael~E}}, \bibinfo {author} {\bibfnamefont {Timothy}\ \bibnamefont
  {O’Leary}}, and\ \bibinfo {author} {\bibfnamefont {Christopher~D}\
  \bibnamefont {Harvey}}} (\bibinfo {year} {2019}),\ \bibfield  {title}
  {\enquote {\bibinfo {title} {Causes and consequences of representational
  drift},}\ }\href {https://doi.org/https://doi.org/10.1016/j.conb.2019.08.005}
  {\bibfield  {journal} {\bibinfo  {journal} {Current Opinion in Neurobiology}\
  }\textbf {\bibinfo {volume} {58}},\ \bibinfo {pages} {141--147}},\ \bibinfo
  {note} {computational Neuroscience}\BibitemShut {NoStop}%
\bibitem [{\citenamefont {Saad}\ and\ \citenamefont
  {Solla}(1995)}]{saad1995line}%
  \BibitemOpen
  \bibfield  {author} {\bibinfo {author} {\bibnamefont {Saad}, \bibfnamefont
  {David}}, and\ \bibinfo {author} {\bibfnamefont {Sara~A}\ \bibnamefont
  {Solla}}} (\bibinfo {year} {1995}),\ \bibfield  {title} {\enquote {\bibinfo
  {title} {On-line learning in soft committee machines},}\ }\href@noop {}
  {\bibfield  {journal} {\bibinfo  {journal} {Physical Review E}\ }\textbf
  {\bibinfo {volume} {52}}~(\bibinfo {number} {4}),\ \bibinfo {pages}
  {4225}}\BibitemShut {NoStop}%
\bibitem [{\citenamefont {Saxe}\ \emph {et~al.}(2013)\citenamefont {Saxe},
  \citenamefont {McClelland},\ and\ \citenamefont {Ganguli}}]{saxe2013exact}%
  \BibitemOpen
  \bibfield  {author} {\bibinfo {author} {\bibnamefont {Saxe}, \bibfnamefont
  {Andrew~M}}, \bibinfo {author} {\bibfnamefont {James~L}\ \bibnamefont
  {McClelland}}, and\ \bibinfo {author} {\bibfnamefont {Surya}\ \bibnamefont
  {Ganguli}}} (\bibinfo {year} {2013}),\ \bibfield  {title} {\enquote {\bibinfo
  {title} {Exact solutions to the nonlinear dynamics of learning in deep linear
  neural networks},}\ }\href@noop {} {\bibfield  {journal} {\bibinfo  {journal}
  {arXiv preprint arXiv:1312.6120}\ }}\Eprint {https://arxiv.org/abs/1312.6120}
  {arXiv:1312.6120} \BibitemShut {NoStop}%
\bibitem [{\citenamefont {Sorscher}\ \emph {et~al.}(2022)\citenamefont
  {Sorscher}, \citenamefont {Ganguli},\ and\ \citenamefont
  {Sompolinsky}}]{sorscher2022fewshot}%
  \BibitemOpen
  \bibfield  {author} {\bibinfo {author} {\bibnamefont {Sorscher},
  \bibfnamefont {Ben}}, \bibinfo {author} {\bibfnamefont {Surya}\ \bibnamefont
  {Ganguli}}, and\ \bibinfo {author} {\bibfnamefont {Haim}\ \bibnamefont
  {Sompolinsky}}} (\bibinfo {year} {2022}),\ \bibfield  {title} {\enquote
  {\bibinfo {title} {Neural representational geometry underlies few-shot
  concept learning},}\ }\href {https://doi.org/10.1073/pnas.2200800119}
  {\bibfield  {journal} {\bibinfo  {journal} {Proceedings of the National
  Academy of Sciences}\ }\textbf {\bibinfo {volume} {119}}~(\bibinfo {number}
  {43}),\ \bibinfo {pages} {e2200800119}},\ \Eprint
  {https://arxiv.org/abs/https://www.pnas.org/doi/pdf/10.1073/pnas.2200800119}
  {https://www.pnas.org/doi/pdf/10.1073/pnas.2200800119} \BibitemShut {NoStop}%
\bibitem [{\citenamefont {Steinmetz}\ \emph {et~al.}(2021)\citenamefont
  {Steinmetz}, \citenamefont {Aydin}, \citenamefont {Lebedeva}, \citenamefont
  {Okun}, \citenamefont {Pachitariu}, \citenamefont {Bauza}, \citenamefont
  {Beau}, \citenamefont {Bhagat}, \citenamefont {Böhm}, \citenamefont {Broux},
  \citenamefont {Chen}, \citenamefont {Colonell}, \citenamefont {Gardner},
  \citenamefont {Karsh}, \citenamefont {Kloosterman}, \citenamefont
  {Kostadinov}, \citenamefont {Mora-Lopez}, \citenamefont {O’Callaghan},
  \citenamefont {Park}, \citenamefont {Putzeys}, \citenamefont {Sauerbrei},
  \citenamefont {van Daal}, \citenamefont {Vollan}, \citenamefont {Wang},
  \citenamefont {Welkenhuysen}, \citenamefont {Ye}, \citenamefont {Dudman},
  \citenamefont {Dutta}, \citenamefont {Hantman}, \citenamefont {Harris},
  \citenamefont {Lee}, \citenamefont {Moser}, \citenamefont {O’Keefe},
  \citenamefont {Renart}, \citenamefont {Svoboda}, \citenamefont {Häusser},
  \citenamefont {Haesler}, \citenamefont {Carandini},\ and\ \citenamefont
  {Harris}}]{steinmetz2021neuropixels}%
  \BibitemOpen
  \bibfield  {author} {\bibinfo {author} {\bibnamefont {Steinmetz},
  \bibfnamefont {Nicholas~A}}, \bibinfo {author} {\bibfnamefont {Cagatay}\
  \bibnamefont {Aydin}}, \bibinfo {author} {\bibfnamefont {Anna}\ \bibnamefont
  {Lebedeva}}, \bibinfo {author} {\bibfnamefont {Michael}\ \bibnamefont
  {Okun}}, \bibinfo {author} {\bibfnamefont {Marius}\ \bibnamefont
  {Pachitariu}}, \bibinfo {author} {\bibfnamefont {Marius}\ \bibnamefont
  {Bauza}}, \bibinfo {author} {\bibfnamefont {Maxime}\ \bibnamefont {Beau}},
  \bibinfo {author} {\bibfnamefont {Jai}\ \bibnamefont {Bhagat}}, \bibinfo
  {author} {\bibfnamefont {Claudia}\ \bibnamefont {Böhm}}, \bibinfo {author}
  {\bibfnamefont {Martijn}\ \bibnamefont {Broux}}, \bibinfo {author}
  {\bibfnamefont {Susu}\ \bibnamefont {Chen}}, \bibinfo {author} {\bibfnamefont
  {Jennifer}\ \bibnamefont {Colonell}}, \bibinfo {author} {\bibfnamefont
  {Richard~J.}\ \bibnamefont {Gardner}}, \bibinfo {author} {\bibfnamefont
  {Bill}\ \bibnamefont {Karsh}}, \bibinfo {author} {\bibfnamefont {Fabian}\
  \bibnamefont {Kloosterman}}, \bibinfo {author} {\bibfnamefont {Dimitar}\
  \bibnamefont {Kostadinov}}, \bibinfo {author} {\bibfnamefont {Carolina}\
  \bibnamefont {Mora-Lopez}}, \bibinfo {author} {\bibfnamefont {John}\
  \bibnamefont {O’Callaghan}}, \bibinfo {author} {\bibfnamefont {Junchol}\
  \bibnamefont {Park}}, \bibinfo {author} {\bibfnamefont {Jan}\ \bibnamefont
  {Putzeys}}, \bibinfo {author} {\bibfnamefont {Britton}\ \bibnamefont
  {Sauerbrei}}, \bibinfo {author} {\bibfnamefont {Rik J.~J.}\ \bibnamefont {van
  Daal}}, \bibinfo {author} {\bibfnamefont {Abraham~Z.}\ \bibnamefont
  {Vollan}}, \bibinfo {author} {\bibfnamefont {Shiwei}\ \bibnamefont {Wang}},
  \bibinfo {author} {\bibfnamefont {Marleen}\ \bibnamefont {Welkenhuysen}},
  \bibinfo {author} {\bibfnamefont {Zhiwen}\ \bibnamefont {Ye}}, \bibinfo
  {author} {\bibfnamefont {Joshua~T.}\ \bibnamefont {Dudman}}, \bibinfo
  {author} {\bibfnamefont {Barundeb}\ \bibnamefont {Dutta}}, \bibinfo {author}
  {\bibfnamefont {Adam~W.}\ \bibnamefont {Hantman}}, \bibinfo {author}
  {\bibfnamefont {Kenneth~D.}\ \bibnamefont {Harris}}, \bibinfo {author}
  {\bibfnamefont {Albert~K.}\ \bibnamefont {Lee}}, \bibinfo {author}
  {\bibfnamefont {Edvard~I.}\ \bibnamefont {Moser}}, \bibinfo {author}
  {\bibfnamefont {John}\ \bibnamefont {O’Keefe}}, \bibinfo {author}
  {\bibfnamefont {Alfonso}\ \bibnamefont {Renart}}, \bibinfo {author}
  {\bibfnamefont {Karel}\ \bibnamefont {Svoboda}}, \bibinfo {author}
  {\bibfnamefont {Michael}\ \bibnamefont {Häusser}}, \bibinfo {author}
  {\bibfnamefont {Sebastian}\ \bibnamefont {Haesler}}, \bibinfo {author}
  {\bibfnamefont {Matteo}\ \bibnamefont {Carandini}}, and\ \bibinfo {author}
  {\bibfnamefont {Timothy~D.}\ \bibnamefont {Harris}}} (\bibinfo {year}
  {2021}),\ \bibfield  {title} {\enquote {\bibinfo {title} {Neuropixels 2.0: A
  miniaturized high-density probe for stable, long-term brain recordings},}\
  }\href {https://doi.org/10.1126/science.abf4588} {\bibfield  {journal}
  {\bibinfo  {journal} {Science}\ }\textbf {\bibinfo {volume} {372}}~(\bibinfo
  {number} {6539}),\ \bibinfo {pages} {eabf4588}}\BibitemShut {NoStop}%
\bibitem [{\citenamefont {Stiso}\ and\ \citenamefont
  {Bassett}(2018)}]{stiso2018spatial}%
  \BibitemOpen
  \bibfield  {author} {\bibinfo {author} {\bibnamefont {Stiso}, \bibfnamefont
  {Jennifer}}, and\ \bibinfo {author} {\bibfnamefont {Dani~S.}\ \bibnamefont
  {Bassett}}} (\bibinfo {year} {2018}),\ \bibfield  {title} {\enquote {\bibinfo
  {title} {Spatial embedding imposes constraints on neuronal network
  architectures},}\ }\href
  {https://doi.org/https://doi.org/10.1016/j.tics.2018.09.007} {\bibfield
  {journal} {\bibinfo  {journal} {Trends in Cognitive Sciences}\ }\textbf
  {\bibinfo {volume} {22}}~(\bibinfo {number} {12}),\ \bibinfo {pages}
  {1127--1142}}\BibitemShut {NoStop}%
\bibitem [{\citenamefont {Sucholutsky}\ \emph {et~al.}(2024)\citenamefont
  {Sucholutsky}, \citenamefont {Muttenthaler}, \citenamefont {Weller},
  \citenamefont {Peng}, \citenamefont {Bobu}, \citenamefont {Kim},
  \citenamefont {Love}, \citenamefont {Cueva}, \citenamefont {Grant},
  \citenamefont {Groen}, \citenamefont {Achterberg}, \citenamefont {Tenenbaum},
  \citenamefont {Collins}, \citenamefont {Hermann}, \citenamefont {Oktar},
  \citenamefont {Greff}, \citenamefont {Hebart}, \citenamefont {Cloos},
  \citenamefont {Kriegeskorte}, \citenamefont {Jacoby}, \citenamefont {Zhang},
  \citenamefont {Marjieh}, \citenamefont {Geirhos}, \citenamefont {Chen},
  \citenamefont {Kornblith}, \citenamefont {Rane}, \citenamefont {Konkle},
  \citenamefont {O'Connell}, \citenamefont {Unterthiner}, \citenamefont
  {Lampinen}, \citenamefont {Müller}, \citenamefont {Toneva},\ and\
  \citenamefont {Griffiths}}]{sucholutsky2024aligned}%
  \BibitemOpen
  \bibfield  {author} {\bibinfo {author} {\bibnamefont {Sucholutsky},
  \bibfnamefont {Ilia}}, \bibinfo {author} {\bibfnamefont {Lukas}\ \bibnamefont
  {Muttenthaler}}, \bibinfo {author} {\bibfnamefont {Adrian}\ \bibnamefont
  {Weller}}, \bibinfo {author} {\bibfnamefont {Andi}\ \bibnamefont {Peng}},
  \bibinfo {author} {\bibfnamefont {Andreea}\ \bibnamefont {Bobu}}, \bibinfo
  {author} {\bibfnamefont {Been}\ \bibnamefont {Kim}}, \bibinfo {author}
  {\bibfnamefont {Bradley~C.}\ \bibnamefont {Love}}, \bibinfo {author}
  {\bibfnamefont {Christopher~J.}\ \bibnamefont {Cueva}}, \bibinfo {author}
  {\bibfnamefont {Erin}\ \bibnamefont {Grant}}, \bibinfo {author}
  {\bibfnamefont {Iris}\ \bibnamefont {Groen}}, \bibinfo {author}
  {\bibfnamefont {Jascha}\ \bibnamefont {Achterberg}}, \bibinfo {author}
  {\bibfnamefont {Joshua~B.}\ \bibnamefont {Tenenbaum}}, \bibinfo {author}
  {\bibfnamefont {Katherine~M.}\ \bibnamefont {Collins}}, \bibinfo {author}
  {\bibfnamefont {Katherine~L.}\ \bibnamefont {Hermann}}, \bibinfo {author}
  {\bibfnamefont {Kerem}\ \bibnamefont {Oktar}}, \bibinfo {author}
  {\bibfnamefont {Klaus}\ \bibnamefont {Greff}}, \bibinfo {author}
  {\bibfnamefont {Martin~N.}\ \bibnamefont {Hebart}}, \bibinfo {author}
  {\bibfnamefont {Nathan}\ \bibnamefont {Cloos}}, \bibinfo {author}
  {\bibfnamefont {Nikolaus}\ \bibnamefont {Kriegeskorte}}, \bibinfo {author}
  {\bibfnamefont {Nori}\ \bibnamefont {Jacoby}}, \bibinfo {author}
  {\bibfnamefont {Qiuyi}\ \bibnamefont {Zhang}}, \bibinfo {author}
  {\bibfnamefont {Raja}\ \bibnamefont {Marjieh}}, \bibinfo {author}
  {\bibfnamefont {Robert}\ \bibnamefont {Geirhos}}, \bibinfo {author}
  {\bibfnamefont {Sherol}\ \bibnamefont {Chen}}, \bibinfo {author}
  {\bibfnamefont {Simon}\ \bibnamefont {Kornblith}}, \bibinfo {author}
  {\bibfnamefont {Sunayana}\ \bibnamefont {Rane}}, \bibinfo {author}
  {\bibfnamefont {Talia}\ \bibnamefont {Konkle}}, \bibinfo {author}
  {\bibfnamefont {Thomas~P.}\ \bibnamefont {O'Connell}}, \bibinfo {author}
  {\bibfnamefont {Thomas}\ \bibnamefont {Unterthiner}}, \bibinfo {author}
  {\bibfnamefont {Andrew~K.}\ \bibnamefont {Lampinen}}, \bibinfo {author}
  {\bibfnamefont {Klaus-Robert}\ \bibnamefont {Müller}}, \bibinfo {author}
  {\bibfnamefont {Mariya}\ \bibnamefont {Toneva}}, and\ \bibinfo {author}
  {\bibfnamefont {Thomas~L.}\ \bibnamefont {Griffiths}}} (\bibinfo {year}
  {2024}),\ \bibfield  {title} {\enquote {\bibinfo {title} {Getting aligned on
  representational alignment},}\ }\href {https://arxiv.org/abs/2310.13018}
  {\bibfield  {journal} {\bibinfo  {journal} {arXiv}\ }}\Eprint
  {https://arxiv.org/abs/2310.13018} {arXiv:2310.13018 [q-bio.NC]} \BibitemShut
  {NoStop}%
\bibitem [{\citenamefont {Sun}\ \emph {et~al.}(2025)\citenamefont {Sun},
  \citenamefont {Winnubst}, \citenamefont {Natrajan}, \citenamefont {Lai},
  \citenamefont {Kajikawa}, \citenamefont {Bast}, \citenamefont {Michaelos},
  \citenamefont {Gattoni}, \citenamefont {Stringer}, \citenamefont
  {Flickinger}, \citenamefont {Fitzgerald},\ and\ \citenamefont
  {Spruston}}]{sun2025orthogonalized}%
  \BibitemOpen
  \bibfield  {author} {\bibinfo {author} {\bibnamefont {Sun}, \bibfnamefont
  {Weinan}}, \bibinfo {author} {\bibfnamefont {Johan}\ \bibnamefont
  {Winnubst}}, \bibinfo {author} {\bibfnamefont {Maanasa}\ \bibnamefont
  {Natrajan}}, \bibinfo {author} {\bibfnamefont {Chongxi}\ \bibnamefont {Lai}},
  \bibinfo {author} {\bibfnamefont {Koichiro}\ \bibnamefont {Kajikawa}},
  \bibinfo {author} {\bibfnamefont {Arco}\ \bibnamefont {Bast}}, \bibinfo
  {author} {\bibfnamefont {Michalis}\ \bibnamefont {Michaelos}}, \bibinfo
  {author} {\bibfnamefont {Rachel}\ \bibnamefont {Gattoni}}, \bibinfo {author}
  {\bibfnamefont {Carsen}\ \bibnamefont {Stringer}}, \bibinfo {author}
  {\bibfnamefont {Daniel}\ \bibnamefont {Flickinger}}, \bibinfo {author}
  {\bibfnamefont {James~E.}\ \bibnamefont {Fitzgerald}}, and\ \bibinfo {author}
  {\bibfnamefont {Nelson}\ \bibnamefont {Spruston}}} (\bibinfo {year} {2025}),\
  \bibfield  {title} {\enquote {\bibinfo {title} {Learning produces an
  orthogonalized state machine in the hippocampus},}\ }\href
  {https://doi.org/10.1038/s41586-024-08548-w} {\bibfield  {journal} {\bibinfo
  {journal} {Nature}\ }\textbf {\bibinfo {volume} {640}}~(\bibinfo {number}
  {8057}),\ \bibinfo {pages} {165--175}}\BibitemShut {NoStop}%
\bibitem [{\citenamefont {Vaidya}\ \emph {et~al.}(2025)\citenamefont {Vaidya},
  \citenamefont {Li}, \citenamefont {Chitwood}, \citenamefont {Li},\ and\
  \citenamefont {Magee}}]{vaidya2025expanding}%
  \BibitemOpen
  \bibfield  {author} {\bibinfo {author} {\bibnamefont {Vaidya}, \bibfnamefont
  {Sachin~P}}, \bibinfo {author} {\bibfnamefont {Guanchun}\ \bibnamefont {Li}},
  \bibinfo {author} {\bibfnamefont {Raymond~A.}\ \bibnamefont {Chitwood}},
  \bibinfo {author} {\bibfnamefont {Yiding}\ \bibnamefont {Li}}, and\ \bibinfo
  {author} {\bibfnamefont {Jeffrey~C.}\ \bibnamefont {Magee}}} (\bibinfo {year}
  {2025}),\ \bibfield  {title} {\enquote {\bibinfo {title} {Formation of an
  expanding memory representation in the hippocampus},}\ }\href
  {https://doi.org/10.1038/s41593-025-01986-3} {\bibfield  {journal} {\bibinfo
  {journal} {Nature Neuroscience}\ }\textbf {\bibinfo {volume} {28}}~(\bibinfo
  {number} {7}),\ \bibinfo {pages} {1510--1518}}\BibitemShut {NoStop}%
\bibitem [{\citenamefont {Watkin}\ \emph {et~al.}(1993)\citenamefont {Watkin},
  \citenamefont {Rau},\ and\ \citenamefont {Biehl}}]{watkin1993rule}%
  \BibitemOpen
  \bibfield  {author} {\bibinfo {author} {\bibnamefont {Watkin}, \bibfnamefont
  {Timothy L~H}}, \bibinfo {author} {\bibfnamefont {Albrecht}\ \bibnamefont
  {Rau}}, and\ \bibinfo {author} {\bibfnamefont {Michael}\ \bibnamefont
  {Biehl}}} (\bibinfo {year} {1993}),\ \bibfield  {title} {\enquote {\bibinfo
  {title} {The statistical mechanics of learning a rule},}\ }\href
  {https://doi.org/10.1103/RevModPhys.65.499} {\bibfield  {journal} {\bibinfo
  {journal} {Rev. Mod. Phys.}\ }\textbf {\bibinfo {volume} {65}},\ \bibinfo
  {pages} {499--556}}\BibitemShut {NoStop}%
\bibitem [{\citenamefont {Williams}(2024)}]{williams2024equivalence}%
  \BibitemOpen
  \bibfield  {author} {\bibinfo {author} {\bibnamefont {Williams},
  \bibfnamefont {Alex~H}}} (\bibinfo {year} {2024}),\ \bibfield  {title}
  {\enquote {\bibinfo {title} {Equivalence between representational similarity
  analysis, centered kernel alignment, and canonical correlations analysis},}\
  }\href {https://doi.org/10.1101/2024.10.23.619871} {\bibfield  {journal}
  {\bibinfo  {journal} {bioRxiv}\ }10.1101/2024.10.23.619871},\ \Eprint
  {https://arxiv.org/abs/https://www.biorxiv.org/content/early/2024/10/24/2024.10.23.619871.full.pdf}
  {https://www.biorxiv.org/content/early/2024/10/24/2024.10.23.619871.full.pdf}
  \BibitemShut {NoStop}%
\bibitem [{\citenamefont {Williams}\ \emph {et~al.}(2021)\citenamefont
  {Williams}, \citenamefont {Kunz}, \citenamefont {Kornblith},\ and\
  \citenamefont {Linderman}}]{williams2021shape}%
  \BibitemOpen
  \bibfield  {author} {\bibinfo {author} {\bibnamefont {Williams},
  \bibfnamefont {Alex~H}}, \bibinfo {author} {\bibfnamefont {Erin}\
  \bibnamefont {Kunz}}, \bibinfo {author} {\bibfnamefont {Simon}\ \bibnamefont
  {Kornblith}}, and\ \bibinfo {author} {\bibfnamefont {Scott}\ \bibnamefont
  {Linderman}}} (\bibinfo {year} {2021}),\ \bibfield  {title} {\enquote
  {\bibinfo {title} {Generalized shape metrics on neural representations},}\
  }in\ \href
  {https://proceedings.neurips.cc/paper_files/paper/2021/file/252a3dbaeb32e7690242ad3b556e626b-Paper.pdf}
  {\emph {\bibinfo {booktitle} {Advances in Neural Information Processing
  Systems}}},\ Vol.~\bibinfo {volume} {34},\ \bibinfo {editor} {edited by\
  \bibinfo {editor} {\bibfnamefont {M.}~\bibnamefont {Ranzato}}, \bibinfo
  {editor} {\bibfnamefont {A.}~\bibnamefont {Beygelzimer}}, \bibinfo {editor}
  {\bibfnamefont {Y.}~\bibnamefont {Dauphin}}, \bibinfo {editor} {\bibfnamefont
  {P.S.}\ \bibnamefont {Liang}}, \ and\ \bibinfo {editor} {\bibfnamefont
  {J.~Wortman}\ \bibnamefont {Vaughan}}}\ (\bibinfo  {publisher} {Curran
  Associates, Inc.})\ pp.\ \bibinfo {pages} {4738--4750}\BibitemShut {NoStop}%
\bibitem [{\citenamefont {Williams}(1996)}]{williams1996infinite}%
  \BibitemOpen
  \bibfield  {author} {\bibinfo {author} {\bibnamefont {Williams},
  \bibfnamefont {Christopher}}} (\bibinfo {year} {1996}),\ \bibfield  {title}
  {\enquote {\bibinfo {title} {Computing with infinite networks},}\ }in\ \href
  {https://proceedings.neurips.cc/paper_files/paper/1996/file/ae5e3ce40e0404a45ecacaaf05e5f735-Paper.pdf}
  {\emph {\bibinfo {booktitle} {Advances in Neural Information Processing
  Systems}}},\ Vol.~\bibinfo {volume} {9},\ \bibinfo {editor} {edited by\
  \bibinfo {editor} {\bibfnamefont {M.C.}\ \bibnamefont {Mozer}}, \bibinfo
  {editor} {\bibfnamefont {M.}~\bibnamefont {Jordan}}, \ and\ \bibinfo {editor}
  {\bibfnamefont {T.}~\bibnamefont {Petsche}}}\ (\bibinfo  {publisher} {MIT
  Press})\BibitemShut {NoStop}%
\bibitem [{\citenamefont {Yang}\ and\ \citenamefont
  {Hu}(2021)}]{yang2021tensor}%
  \BibitemOpen
  \bibfield  {author} {\bibinfo {author} {\bibnamefont {Yang}, \bibfnamefont
  {Greg}}, and\ \bibinfo {author} {\bibfnamefont {Edward~J}\ \bibnamefont
  {Hu}}} (\bibinfo {year} {2021}),\ \bibfield  {title} {\enquote {\bibinfo
  {title} {Tensor {P}rograms {IV}: Feature learning in infinite-width neural
  networks},}\ }in\ \href@noop {} {\emph {\bibinfo {booktitle} {International
  Conference on Machine Learning}}}\ (\bibinfo {organization} {PMLR})\ pp.\
  \bibinfo {pages} {11727--11737}\BibitemShut {NoStop}%
\bibitem [{\citenamefont {Zavatone-Veth}\ \emph
  {et~al.}(2022{\natexlab{a}})\citenamefont {Zavatone-Veth}, \citenamefont
  {Canatar}, \citenamefont {Ruben},\ and\ \citenamefont
  {Pehlevan}}]{zv2022asymptotics}%
  \BibitemOpen
  \bibfield  {author} {\bibinfo {author} {\bibnamefont {Zavatone-Veth},
  \bibfnamefont {Jacob~A}}, \bibinfo {author} {\bibfnamefont {Abdulkadir}\
  \bibnamefont {Canatar}}, \bibinfo {author} {\bibfnamefont {Benjamin~S}\
  \bibnamefont {Ruben}}, and\ \bibinfo {author} {\bibfnamefont {Cengiz}\
  \bibnamefont {Pehlevan}}} (\bibinfo {year} {2022}{\natexlab{a}}),\ \bibfield
  {title} {\enquote {\bibinfo {title} {Asymptotics of representation learning
  in finite {B}ayesian neural networks},}\ }\href
  {https://doi.org/10.1088/1742-5468/ac98a6} {\bibfield  {journal} {\bibinfo
  {journal} {Journal of Statistical Mechanics: Theory and Experiment}\ }\textbf
  {\bibinfo {volume} {2022}}~(\bibinfo {number} {11}),\ \bibinfo {pages}
  {114008}}\BibitemShut {NoStop}%
\bibitem [{\citenamefont {Zavatone-Veth}\ and\ \citenamefont
  {Pehlevan}(2021)}]{zv2021scale}%
  \BibitemOpen
  \bibfield  {author} {\bibinfo {author} {\bibnamefont {Zavatone-Veth},
  \bibfnamefont {Jacob~A}}, and\ \bibinfo {author} {\bibfnamefont {Cengiz}\
  \bibnamefont {Pehlevan}}} (\bibinfo {year} {2021}),\ \bibfield  {title}
  {\enquote {\bibinfo {title} {Depth induces scale-averaging in
  overparameterized linear {Bayesian} neural networks},}\ }in\ \href
  {https://doi.org/10.1109/IEEECONF53345.2021.9723137} {\emph {\bibinfo
  {booktitle} {Asilomar Conference on Signals, Systems, and Computers}}},\
  Vol.~\bibinfo {volume} {55},\ \Eprint {https://arxiv.org/abs/2111.11954}
  {arXiv:2111.11954} \BibitemShut {NoStop}%
\bibitem [{\citenamefont {Zavatone-Veth}\ and\ \citenamefont
  {Pehlevan}(2022)}]{zv2022capacity}%
  \BibitemOpen
  \bibfield  {author} {\bibinfo {author} {\bibnamefont {Zavatone-Veth},
  \bibfnamefont {Jacob~A}}, and\ \bibinfo {author} {\bibfnamefont {Cengiz}\
  \bibnamefont {Pehlevan}}} (\bibinfo {year} {2022}),\ \bibfield  {title}
  {\enquote {\bibinfo {title} {On neural network kernels and the storage
  capacity problem},}\ }\href {https://doi.org/10.1162/neco_a_01494} {\bibfield
   {journal} {\bibinfo  {journal} {Neural Computation}\ }\textbf {\bibinfo
  {volume} {34}}~(\bibinfo {number} {5}),\ \bibinfo {pages} {1136--1142}},\
  \Eprint {https://arxiv.org/abs/2201.04669} {arXiv:2201.04669} \BibitemShut
  {NoStop}%
\bibitem [{\citenamefont {Zavatone-Veth}\ \emph
  {et~al.}(2022{\natexlab{b}})\citenamefont {Zavatone-Veth}, \citenamefont
  {Tong},\ and\ \citenamefont {Pehlevan}}]{zavatone2022contrasting}%
  \BibitemOpen
  \bibfield  {author} {\bibinfo {author} {\bibnamefont {Zavatone-Veth},
  \bibfnamefont {Jacob~A}}, \bibinfo {author} {\bibfnamefont {William~L}\
  \bibnamefont {Tong}}, and\ \bibinfo {author} {\bibfnamefont {Cengiz}\
  \bibnamefont {Pehlevan}}} (\bibinfo {year} {2022}{\natexlab{b}}),\ \bibfield
  {title} {\enquote {\bibinfo {title} {Contrasting random and learned features
  in deep {B}ayesian linear regression},}\ }\href@noop {} {\bibfield  {journal}
  {\bibinfo  {journal} {Physical Review E}\ }\textbf {\bibinfo {volume}
  {105}}~(\bibinfo {number} {6}),\ \bibinfo {pages} {064118}}\BibitemShut
  {NoStop}%
\bibitem [{\citenamefont {Zdeborov{\'a}}\ and\ \citenamefont
  {Krzakala}(2016)}]{zdeborova2016thresholds}%
  \BibitemOpen
  \bibfield  {author} {\bibinfo {author} {\bibnamefont {Zdeborov{\'a}},
  \bibfnamefont {Lenka}}, and\ \bibinfo {author} {\bibfnamefont {Florent}\
  \bibnamefont {Krzakala}}} (\bibinfo {year} {2016}),\ \bibfield  {title}
  {\enquote {\bibinfo {title} {Statistical physics of inference: thresholds and
  algorithms},}\ }\href {https://doi.org/10.1080/00018732.2016.1211393}
  {\bibfield  {journal} {\bibinfo  {journal} {Advances in Physics}\ }\textbf
  {\bibinfo {volume} {65}}~(\bibinfo {number} {5}),\ \bibinfo {pages}
  {453--552}}\BibitemShut {NoStop}%
\bibitem [{\citenamefont {Zhang}\ \emph {et~al.}(2023)\citenamefont {Zhang},
  \citenamefont {Pan}, \citenamefont {Jung}, \citenamefont {Halpern},
  \citenamefont {Eichhorn}, \citenamefont {Lei}, \citenamefont {Cohen},
  \citenamefont {Smith}, \citenamefont {Tasic}, \citenamefont {Yao},
  \citenamefont {Zeng},\ and\ \citenamefont {Zhuang}}]{zhang2023atlas}%
  \BibitemOpen
  \bibfield  {author} {\bibinfo {author} {\bibnamefont {Zhang}, \bibfnamefont
  {Meng}}, \bibinfo {author} {\bibfnamefont {Xingjie}\ \bibnamefont {Pan}},
  \bibinfo {author} {\bibfnamefont {Won}\ \bibnamefont {Jung}}, \bibinfo
  {author} {\bibfnamefont {Aaron~R.}\ \bibnamefont {Halpern}}, \bibinfo
  {author} {\bibfnamefont {Stephen~W.}\ \bibnamefont {Eichhorn}}, \bibinfo
  {author} {\bibfnamefont {Zhiyun}\ \bibnamefont {Lei}}, \bibinfo {author}
  {\bibfnamefont {Limor}\ \bibnamefont {Cohen}}, \bibinfo {author}
  {\bibfnamefont {Kimberly~A.}\ \bibnamefont {Smith}}, \bibinfo {author}
  {\bibfnamefont {Bosiljka}\ \bibnamefont {Tasic}}, \bibinfo {author}
  {\bibfnamefont {Zizhen}\ \bibnamefont {Yao}}, \bibinfo {author}
  {\bibfnamefont {Hongkui}\ \bibnamefont {Zeng}}, and\ \bibinfo {author}
  {\bibfnamefont {Xiaowei}\ \bibnamefont {Zhuang}}} (\bibinfo {year} {2023}),\
  \bibfield  {title} {\enquote {\bibinfo {title} {Molecularly defined and
  spatially resolved cell atlas of the whole mouse brain},}\ }\href
  {https://doi.org/10.1038/s41586-023-06808-9} {\bibfield  {journal} {\bibinfo
  {journal} {Nature}\ }\textbf {\bibinfo {volume} {624}}~(\bibinfo {number}
  {7991}),\ \bibinfo {pages} {343--354}}\BibitemShut {NoStop}%
\bibitem [{\citenamefont {Zhong}\ \emph {et~al.}(2025)\citenamefont {Zhong},
  \citenamefont {Baptista}, \citenamefont {Gattoni}, \citenamefont {Arnold},
  \citenamefont {Flickinger}, \citenamefont {Stringer},\ and\ \citenamefont
  {Pachitariu}}]{zhong2025unsupervised}%
  \BibitemOpen
  \bibfield  {author} {\bibinfo {author} {\bibnamefont {Zhong}, \bibfnamefont
  {Lin}}, \bibinfo {author} {\bibfnamefont {Scott}\ \bibnamefont {Baptista}},
  \bibinfo {author} {\bibfnamefont {Rachel}\ \bibnamefont {Gattoni}}, \bibinfo
  {author} {\bibfnamefont {Jon}\ \bibnamefont {Arnold}}, \bibinfo {author}
  {\bibfnamefont {Daniel}\ \bibnamefont {Flickinger}}, \bibinfo {author}
  {\bibfnamefont {Carsen}\ \bibnamefont {Stringer}}, and\ \bibinfo {author}
  {\bibfnamefont {Marius}\ \bibnamefont {Pachitariu}}} (\bibinfo {year}
  {2025}),\ \bibfield  {title} {\enquote {\bibinfo {title} {Unsupervised
  pretraining in biological neural networks},}\ }\href
  {https://doi.org/10.1038/s41586-025-09180-y} {\bibfield  {journal} {\bibinfo
  {journal} {Nature}\ }10.1038/s41586-025-09180-y}\BibitemShut {NoStop}%
\end{thebibliography}%

\end{document}